\documentclass[twocolumn,epjc3]{svjour3}          

\RequirePackage[T1]{fontenc}

\smartqed  

\usepackage{amssymb}
\usepackage{newtxtext}
\usepackage{newtxmath}
\RequirePackage{graphicx}
\RequirePackage{mathptmx}      
\RequirePackage{flushend}
\RequirePackage[numbers,sort&compress]{natbib}
\RequirePackage[colorlinks,citecolor=blue,urlcolor=blue,linkcolor=blue]{hyperref}

\journalname{Eur. Phys. J. C}

\begin{document}

\title{Intrinsic mirror noise in Fabry-Perot based polarimeters: the case for the measurement of vacuum magnetic birefringence}

\author{G.~Zavattini\thanksref{e1,addr1}
        \and
        F.~Della~Valle\thanksref{addr2}
        \and
        A.~Ejlli\thanksref{addr1}
        \and
        W.-T.~Ni\thanksref{addr3}
        \and
        U.~Gastaldi\thanksref{addr1}
        \and
        E.~Milotti\thanksref{addr4}
        \and
        R.~Pengo\thanksref{addr5}
        \and
        G.~Ruoso\thanksref{addr5}
}

\thankstext{e1}{e-mail: guido.zavattini@unife.it}

\institute{INFN, Sez. di Ferrara and Dip. di Fisica e Scienze della Terra, Universit\`a di Ferrara, via G. Saragat 1, Edificio C, I-44122 Ferrara (FE), Italy\label{addr1}
          \and
          INFN, Sez. di Pisa, gruppo collegato di Siena and Dip. di Scienze Fisiche, della Terra e dell'Ambiente, Universit\`a di Siena, via Roma 56, I-53100 Siena (SI), Italy\label{addr2}
          \and
          Department of Physics, National Tsing Hua University, Hsinchu, Taiwan, 30013 Republic of China \label{addr3}
          \and
          INFN, Sez. di Trieste and Dip. di Fisica, Universit\`a di Trieste, via A. Valerio 2, I-34127 Trieste (TS), Italy\label{addr4}
          \and
          INFN, Lab. Naz. di Legnaro, viale dell'Universit\`a 2, I-35020 Legnaro (PD), Italy\label{addr5}
}

\date{Received: date / Accepted: date}

\maketitle

\begin{abstract}
Although experimental efforts have been active for about 30 years now, a direct laboratory observation of vacuum magnetic birefringence, an effect due to vacuum fluctuations, still needs confirmation. Indeed, the predicted birefringence of vacuum is $\Delta n = 4.0\times 10^{-24}$ @ 1~T. One of the key ingredients when designing a polarimeter capable of detecting such a small birefringence is a long optical path length within the magnetic field and a time dependent effect. To lengthen the optical path within the magnetic field a Fabry-Perot optical cavity is generally used with a finesse ranging from ${\cal F} \approx 10^4$ to ${\cal F} \approx7\times 10^5$. Interestingly, there is a difficulty in reaching the predicted shot noise limit of such polarimeters. 
We have measured the ellipticity and rotation noises along with a Cotton-Mouton and a Faraday effect as a function of the finesse of the cavity of the PVLAS polarimeter. The observations are consistent with the idea that the cavity mirrors generate a birefringence-dominated noise whose ellipticity is amplified by the cavity itself. The optical path difference sensitivity at $10\;$Hz is $S_{\Delta{\cal D}}=6\times 10^{-19}\;$m$/\sqrt{\rm Hz}$, a value which we believe is consistent with an intrinsic thermal noise in the mirror coatings.
\end{abstract}


\section{Introduction}

The development of extremely sensitive polarimeters has been driven in recent years by attempts to measure directly vacuum magnetic birefringence, a non linear quantum electrodynamic effect in vacuum closely related to light-by-light elastic scattering. Non linear electrodynamic effects in vacuum were first predicted in 1935 by the Euler-Kockel perturbative effective Lagrangian density \cite{EK,Eu,HE,W,KN,S,Dunne,Er,BB1,BB2,BB3,Ad},
\begin{eqnarray}
\label{eq:EK}
{\cal L}_{\rm EK}&=&\frac{1}{2\mu_{\rm 0}}\left(\frac{E^{2}}{c^{2}}-B^{2}\right)+\nonumber\\
&+&\frac{A_{e}}{\mu_{\rm 0}}\left[\left(\frac{E^{2}}{c^{2}}-B^{2}\right)^{2}+7\left(\frac{\vec{E}}{c}\cdot\vec{B}\right)^{2}\right]
\end{eqnarray}
which takes into account vacuum fluctuations with the creation of electron-positron pairs. As of today, ${\cal L}_{\rm EK}$ still needs direct experimental confirmation at low energies. This Lagrangian density is valid for field intensities much lower than the critical values: $B \ll B_{\rm crit}={m_e^{2}c^{2}}/{e \hbar}=4.4\times 10^{9}$~T, $E \ll E_{\rm crit}={m_e^{2}c^{3}}/{e \hbar}=1.3\times 10^{18}$~V/m. Here
 \begin{equation}
 \label{eq:Ae}
A_e=\frac{2}{45\mu_{0}}\frac{\alpha^2 \mathchar'26\mkern-10mu\lambda_e^{3}}{m_{e}c^{2}}= \frac{\alpha}{90\pi}\frac{1}{B_{\rm crit}^2}=1.32\times 10^{-24} {\rm{~T}}^{-2}
\end{equation}
describes the entity of the quantum correction to Classical Electrodynamics. The Lagrangian density (\ref{eq:EK}) predicts that vacuum becomes birefringent in the presence of either an external electric or magnetic field \cite{Er,BB1,BB2,BB3,Ad}. In the case of an external magnetic field the unitary birefringence, to order $\alpha^2$, is expected to be
\begin{equation}
\frac{\Delta n}{B^2} =3A_e = \frac{2}{15\mu_{0}}\frac{\alpha^2 \mathchar'26\mkern-10mu\lambda_e^{3}}{m_{e}c^{2}}=3.96\times10^{-24} {\rm{~T}}^{-2}.
\label{eq:3Ae}
\end{equation}
In the presence of an external electric field, $B^2$ is replaced by $-\left(E/c\right)^2$.

Due to this birefringence, a linearly polarised beam of light propagating perpendicularly to the external magnetic field acquires an ellipticity $\psi$
\[
\psi=\psi_0\sin2\vartheta=\pi\frac{\int_0^L\Delta n\;dl }{\lambda}\sin2\vartheta=\pi\frac{3A_e \int_0^L B^2 \;dl}{\lambda}\sin2\vartheta
\]
where $\psi_0$ is the ellipticity amplitude, $\lambda$ is the wavelength of the light, $L$ is the length of the magnetic field and $\vartheta$ is the angle between the magnetic field and the polarisation direction. 
With the parameters of the PVLAS experiment \cite{DellaValle2014}, $B = 2.5\;$T, $L = 1.64\;$m and $\lambda = 1064\;$nm, the induced ellipticity is $\psi = 1.2\times 10^{-17}$, an extremely small value. As we will see in the following section, one way to enhance the induced ellipticity is to increase the effective length of the magnetic field region using a Fabry-Perot cavity with finesse ${\cal F}$. Such a cavity enhances an ellipticity (or a rotation) by a factor $N = 2{\cal F}/\pi$ \cite{Rosenberg1964,Pace1994,Jacob1995,Zavattini2006} which, today, can be as high as $N = 4.5\times 10^5$ \cite{OE2014}.

Several experiments are underway, of which the most sensitive at present are based on polarimeters with such very high finesse Fabry-Perot cavities \cite{DellaValle2015,Q&A2010,BMV2014,OVAL}. Furthermore all of these experiments use variable magnetic fields in order to induce a time dependent effect hence further increasing their sensitivities. This time dependence can be obtained either by varying the magnetic field intensity, in which case $\Delta n = \Delta n(t)$, or by rotating the field direction in a plane perpendicular to the propagation direction such that $\vartheta = \vartheta(t)$. In this second case, adopted by the PVLAS experiment \cite{DellaValle2015} with $N = 4.5\times 10^5$, the signal to be measured is 
\begin{eqnarray}\nonumber
\label{eq:Psi}
\Psi(t)=N\psi(t) &=& N\pi\frac{3A_e \int_0^L B^2 \;dl}{\lambda}\sin2\vartheta(t) \\&=& 5\times 10^{-11}\sin2\vartheta(t)\nonumber
\end{eqnarray}
At present the lowest measured value for $\Delta n/B^2$ is \cite{Aldo}
\[
\Delta n^{\rm PVLAS}/B^2 = (1.9\pm 2.7)\times 10^{-23}\;{\rm T}^{-2}.
\]
The experimental uncertainty on this value is a factor of about seven above the predicted QED value in equation (\ref{eq:3Ae}).
\section{Polarimetry: state of the art}
\begin{figure}[htb]
\begin{center}
\includegraphics[width=8.5cm]{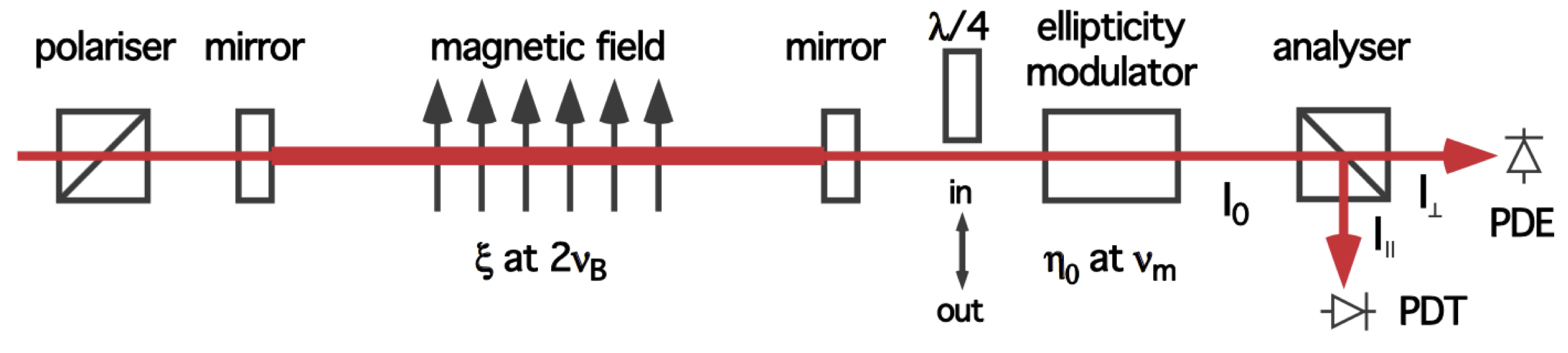}
\end{center}
\caption{A polarimeter based on a Fabry-Perot cavity with a time-dependent signal and heterodyne detection. PDE: Extinction Photodiode; PDT: Transmission Photodiode.}
\label{heterodyne}
\end{figure}

A scheme of the PVLAS polarimeter is shown in figure \ref{heterodyne}. A beam first passes through a polariser and then enters the Fabry-Perot cavity composed of two high-reflectance mirrors placed at a distance $D = 3.303\;$m apart. Between the mirrors is a magnetic field of length $L$ which, in the case of the PVLAS experiment, is generated by two identical rotating permanent magnets characterised by the total parameter $\int_0^L{B^2 dl} = 10.25\;$T$^2$m resulting in an average field $B = 2.5\;$T over a length $L = 1.64\;$m. These two magnets have been rotated up to a frequency $\nu_{B} = 23\;$Hz. Given the dependence of the induced ellipticity $\psi(t)$ with $2\vartheta(t)$, the ellipticity signal due to magnetic birefringence has a frequency component at $2\nu_{B}$. Since the magnetic field could in principle also generate rotations $\phi(t)$ due to a magnetic dichroism (for example from axion-like particles \cite{MPZ}) in figure \ref{heterodyne} the total effect is indicated with a complex number $\xi = \phi + i\psi$. Indeed one can assign an absolute phase to the electric field of the light such that a rotation is described by a pure real number whereas an ellipticity is a pure imaginary quantity. After the output mirror, an ellipticity modulator adds a known ellipticity of amplitude $\eta_0$ to the polarisation at a frequency $\nu_{m}$. The beam 
of power $I_0$ then passes through an analyser which divides the light into two polarisation components: parallel and perpendicular to the input polariser, $I_\parallel$ and $I_\perp$ respectively. These beams are collected by the two photodiodes PDT and PDE with efficiencies $q = 0.7\;$A/W.

If all ellipticities (and rotations) are small these add algebraically. In the presence of both a rotation $\Phi(t)=N\phi(t)$ and an ellipticity $\Psi(t) = N\psi(t)$ and 
without the presence of the quarter-wave plate shown in figure \ref{heterodyne}, the power reaching PDE is
\begin{eqnarray}
I^{\rm ell}_\perp(t)&=& \int\limits_{beam}\epsilon_0c|E_\perp(t)|^2 d\Sigma \simeq I_0 \left|i\eta(t)+i\Psi(t)+\Phi(t)\right|^2 =\nonumber\\ &=&I_0\left[\eta^2(t)+2\eta(t)\Psi(t)+\Phi(t)^2+\Psi(t)^2\right].
\end{eqnarray}
As can be seen, only the ellipticity $\Psi(t)$ beats with the effect of the modulator $\eta(t)$. If $\eta(t)$ and $\Psi(t)$ are sinusoidal functions at frequencies $\nu_m$ and $v$ respectively, the product $2\eta(t)\Psi(t)$ generates Fourier components at $\nu_m\pm\nu$.

By demodulating the current signal from PDE $i^{\rm ell}_\perp(t) = qI^{\rm ell}_\perp(t)$ at the frequencies $\nu_{m}$ and $2\nu_{m}$, one obtains the in-phase Fourier components
\[
i_{\nu_{m}}(\nu) = 2I_0q\eta_0\Psi_0(\nu) \quad {\rm and}\quad i_{2\nu_{m}}({\rm dc})=qI_0\eta_0^2/2
\]
 from which one can extract the amplitude (with relative sign) for $\Psi_0$:
\begin{equation}
\label{eq:ell}
\Psi_0(\nu)= \frac{i_{\nu_{m}}(\nu)}{2qI_0\eta_0}=\frac{i_{\nu_{m}}(\nu)}{2\sqrt{2qI_\parallel i_{2\nu_{m}}({\rm dc})}}=\frac{\eta_0}{4}\frac{i_{\nu_{m}}(\nu)}{i_{2\nu_{m}}({\rm dc})}.
\end{equation}

By inserting the quarter-wave plate with one of its axes aligned with the polarisation, the ellipticity generated by the magnetic field becomes a rotation and vice-versa \cite{Bregant2008}. In this case, the power reaching PDE is
\begin{eqnarray}
I^{\rm rot}_\perp(t)&=& \int\limits_{beam}\epsilon_0c|E_\perp(t)|^2  d\Sigma\simeq I_0\left|i\eta(t)\pm\Psi(t)\mp i\Phi(t)\right|^2 =\nonumber\\ &=&I_0\left[\eta^2(t)\pm2\eta(t)\Phi(t)+\Phi(t)^2+\Psi(t)^2\right]
\end{eqnarray}
where the signs depend on whether the polarisation is aligned with the fast or the slow axis of the $\lambda/4$ wave plate. 
Again the value of $\Phi_0(\nu)$ can be extracted using the same expressions in equation (\ref{eq:ell}).

In the spectra obtained from equation (\ref{eq:ell}), an ellipticity generated by a magnetic birefringence or a rotation generated by a magnetic dichroism will appear at $\nu = 2\nu_{B}$ whereas a rotation due to a time dependent Faraday effect 
at $\nu_F$ will appear at $\nu = \nu_{F}$.

Given the scheme in figure \ref{heterodyne}, one can determine the expected \emph{peak} ellipticity sensitivity $S_\Psi(\nu)$ of the polarimeter in the presence of various noise sources. All noise contributions will be expressed as electric currents. In general the rms noise measured at the output of the demodulator at a frequency $\nu$ is the incoherent sum of the rms noise densities $S_{+}$ and $S_{-}$ respectively at the frequencies $\nu_m+\nu$ and $\nu_m-\nu$. Generally $|S_{+}|=|S_{-}|=S_{\nu}$. Using equation (\ref{eq:ell}) one finds
\begin{equation}
\label{eq:sens}
S_\Psi(\nu)= \frac{\sqrt 2\sqrt{S_{+}^2+S_{-}^2}}{2qI_0\eta_0}= \frac{S_{\nu}}{qI_0\eta_0}.
\end{equation}
The ultimate \emph{peak} sensitivity $S_\Psi$ of such a polarimeter is given by the shot-noise limit.
The rms current spectral density $i^{\rm shot}$ at PDE due to an incident d.c. light power $I_\perp{\rm (dc)}$ is
\[
i^{\rm shot} = \sqrt{2eqI_\perp{\rm (dc)}},
\]
constant over the whole spectrum. Equation (\ref{eq:sens}) then leads to
\begin{equation}
S^{\rm shot}_\Psi(\nu)=  \sqrt{\frac{2e}{qI_\parallel}\frac{\left(\eta_0^2/2+\sigma^2\right)}{\eta_0^2}}
\end{equation}
where we have introduced the extinction ratio of the polarisers $\sigma^2$ and we have introduced $I_\parallel$ as a measurement of $I_0$. If the modulation amplitude is $\eta_0^2/2\gg\sigma^2$, the above expression simplifies to
\begin{equation}
S^{\rm shot}_\Psi(\nu)=  \sqrt{\frac{e}{qI_\parallel}}.
\label{eq:shot}
\end{equation}
{As will be discussed below, the value of $I_\parallel$ used in the PVLAS setup during the measurements presented in this work is $I_\parallel = 0.7\;$mW from which one obtains a shot noise peak sensitivity of}
\[
S^{\rm shot} _\Psi(\nu)= 1.8\times 10^{-8}\;1/\sqrt{\rm Hz}.
\]
With the effect to be measured $\Psi=5.4\times10^{-11}$ and with the above shot noise the measurement time for a unitary signal-to-noise ratio should be $T = \left({S^{\rm shot} _\Psi}/{\Psi}\right)^2 = 1.1\times 10^5\;$s, in principle a reasonable integration time.

\begin{figure}[htb]
\label{fig:sens}
\begin{center}
\includegraphics[width=8.0cm]{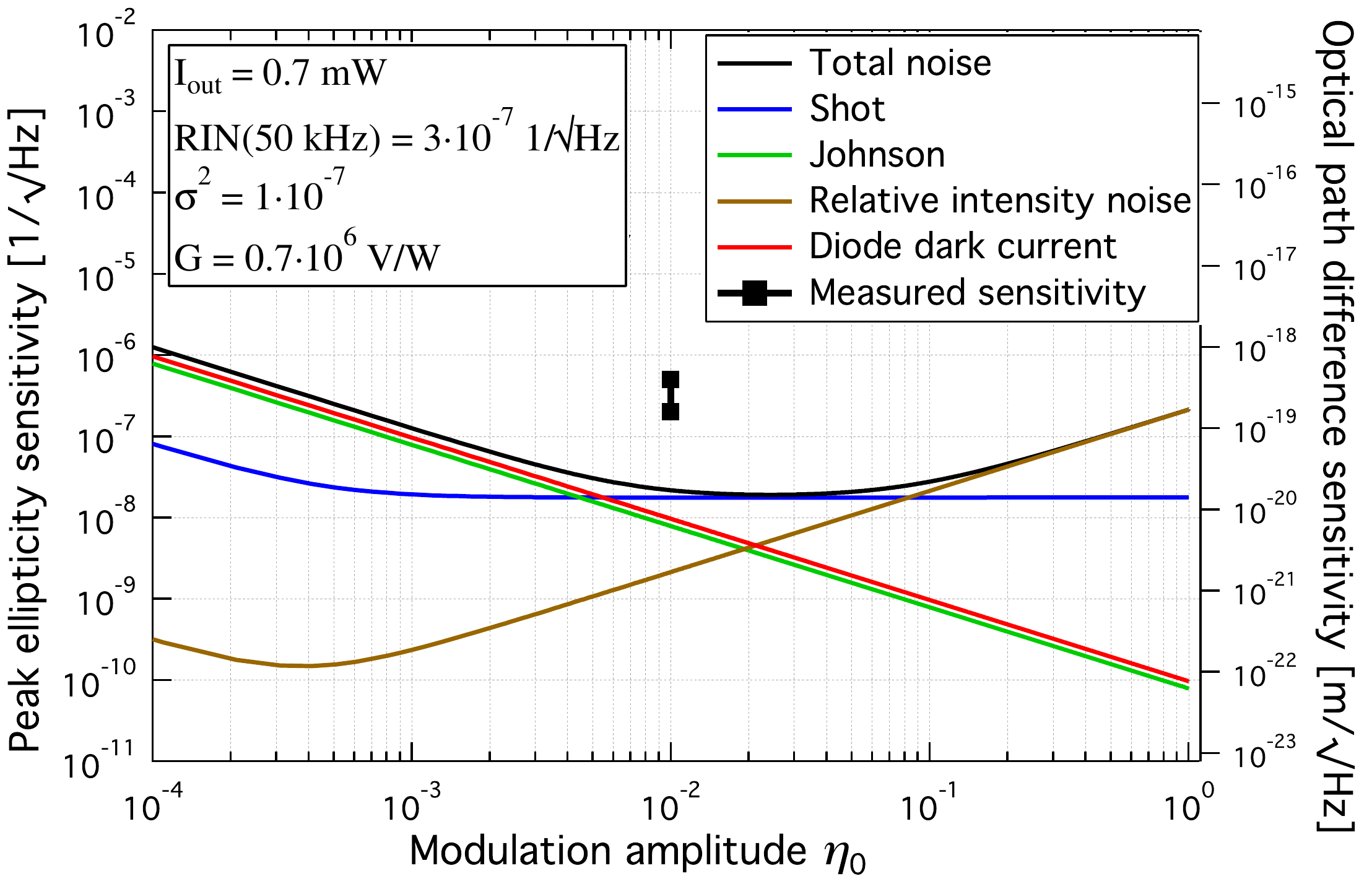}
\end{center}
\caption{Noise budget of the principal noise sources as a function of the modulation amplitude $\eta_0$ for the PVLAS polarimeter. A minimum which coincides with shot-noise sensitivity exists. Superimposed on the plot is the experimental sensitivity between 10~Hz and 20~Hz with $\eta_0 = 10^{-2}$.}
\end{figure}

Considering other known noise sources such as the Johnson noise, $i^{\rm JN}$, the diode dark current noise $i^{\rm DN}$ and the laser's relative intensity noise $i^{\rm RIN}$ one obtains the curves shown in figure \ref{fig:sens} for the sensitivity as a function of the modulation amplitude $\eta_0$ \cite{DellaValle2015}.
As can be seen there is a region in the modulation amplitude around $\eta_0\approx 2\times 10^{-2}$ which should in principle allow shot noise sensitivity.

The out-of-phase quadrature signal  $i^{\rm qu}_{\nu_{m}}(\nu)$ at the output of the demodulator can be used as a good measurement of noise contributions uncorrelated to ellipticity noise. In principle, if $S_\Psi(\nu)$ was limited by one of the wide band noises reported in figure \ref{fig:sens} then $i_{\nu_{m}}(\nu) = i^{\rm qu}_{\nu_{m}}(\nu)$.
Unfortunately this is not the case and the measured sensitivity of the PVLAS polarimeter when measuring ellipticities with $N = 4.5\times 10^5$ is significantly worse than the values in figure \ref{fig:sens}: $S_\Psi^{\rm PVLAS} \sim 3-5\times 10^{-7}\;1/\sqrt{\rm Hz}$ for frequencies $\nu \sim 10 - 20\;{\rm Hz}$ and $\eta_0 = 10^{-2}$. 

To better understand the actual sensitivity of the PVLAS experiment one should consider, rather than the ellipticity, the sensitivity in optical path difference $\Delta{\cal D} =\int\limits_{path}{\Delta n\; dl}$:
\begin{equation}
S_{\Delta{\cal D}} = S_\Psi^{\rm PVLAS}\frac{\lambda}{\pi N}\sim 3 - 6\times 10^{-19}\;{\rm m}/\sqrt{\rm Hz}
\end{equation}
between 10~Hz and 20~Hz.
This value can be compared to the ones for gravitational wave detection using interferometer techniques \cite{virgoqed}. Indeed the quantities to be compared are
\[
S_\Psi^{\rm PVLAS}\frac{\lambda}{2\pi N}\quad\Longleftrightarrow\quad h_{\rm sens} l_{\rm arm}
\]
where $l_{\rm arm}$ is the arm length of the gravitational wave interferometer and $h_{\rm sens}$ is its sensitivity in strain \cite{virgoqed}.
For example in Advanced LIGO \cite{ligo}, with $l_{\rm arm} = 4000\;$m and $h_{\rm sens} \sim 10^{-22}\;1/\sqrt{\rm Hz}$ @ 10 Hz, one finds $S_{\Delta{\cal D}} \sim 4\times 10^{-19}\;$m$/\sqrt{\rm Hz}$, a value similar to the one of the PVLAS experiment. It must also be noted that gravitational wave interferometers are not shot noise limited at these frequencies but are limited by thermal noise of the suspensions and of the mirrors.

\begin{figure}[tb]
\begin{center}
\includegraphics[width=8.5cm]{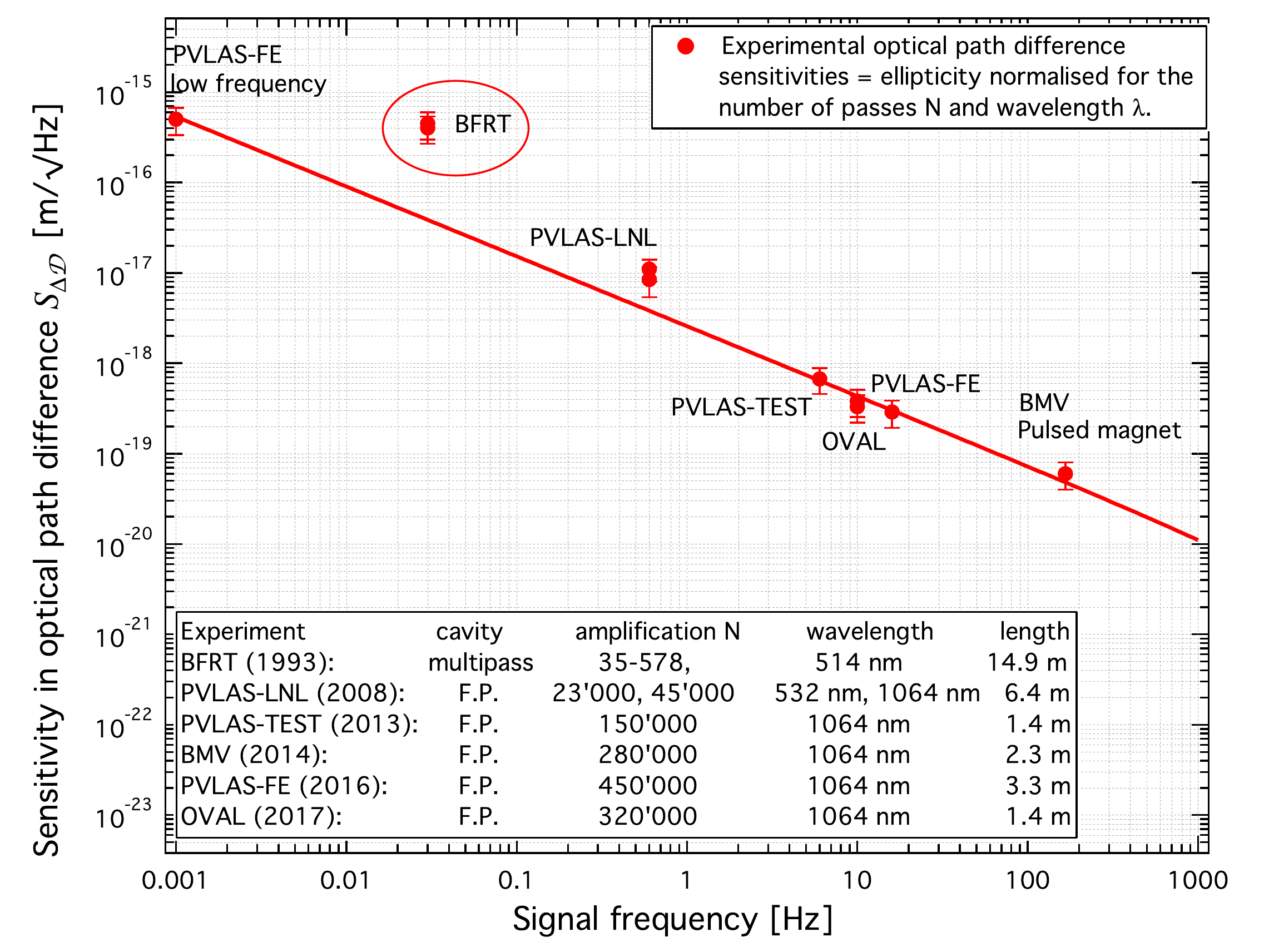}
\end{center}
\caption{Measured optical path difference sensitivity for past and present experiments as a function of their typical working frequency. BFRT \cite{Cameron1993}, PVLAS-LNL \cite{ Bregant2008,Bregant12008}, PVLAS-TEST \cite{DellaValle2013}, BMV \cite{BMV2014}, PVLAS-FE \cite{DellaValle2015}, OVAL \cite{OVAL}. The line is a fit with a power law having excluded the BFRT values. The resulting power is $(-0.78\pm0.03)$.}
\label{fig:path}
\end{figure}

Interestingly, all of the past and present experimental efforts also 
have been limited by a yet to be understood wide band noise. In figure \ref{fig:path} we report the optical path difference sensitivities of past and present experiments dedicated to measuring vacuum magnetic birefringence with optical techniques. These sensitivities are plotted as a function of the frequencies at which each experiment typically works/worked at.
%
%
Although each experiment is characterised by a different finesse of the cavity and uses different detection schemes (heterodyne, homodyne), the sensitivities lie on a common power law $\propto\nu^{x}$ with $x = -0.78\pm0.03$. The only experiment significantly above this common curve is BFRT \cite{Cameron1993}. This is the oldest effort and used a multi-pass cavity with separate optical benches rather than a Fabry-Perot. The mirrors were also of different fabrication. Furthermore all of these sensitivities are well above their expected shot noise limit with the exception of the OVAL experiment which uses a very low power of $10\;\mu $W at the output of the cavity and whose sensitivity coincides with its expected one \cite{OVAL}.

Finally, without the presence of the Fabry-Perot cavity the PVLAS polarimeter reaches shot-noise sensitivity above $\nu \sim 10\;$Hz. Below this frequency the noise is due to pointing fluctuations of the laser beam coupled to birefringence gradients present in all optical elements.
 
A possibile interpretation of the general behaviour shown in figure \ref{fig:path} is that there is an intrinsic birefringence noise being generated in the mirror reflective coatings. Given the order of magnitude of the sensitivities in figure \ref{fig:path} we believe that we have reached a thermal intrinsic noise in birefringence, not induced by the laser power, but due to the mirrors in thermal equilibrium at T $\approx$ 300 K. 
%

To verify whether the excess noise present in the PVLAS experiment is indeed a birefringence noise originating from the reflective coatings of the mirrors we performed a series of measurements both in ellipticity and in rotation as a function of the finesse of our cavity. The measurements were performed with a pure birefringence signal due to Argon gas at a pressure $P\approx 0.85\;$mbar and with an external solenoid generating a Faraday rotation on the input mirror of the cavity. In this paper the results of these measurements are presented showing that indeed the excess noise is dominated by birefringence noise and that the ellipticity noise is proportional to the finesse of the cavity. 

\section{Method}
The basic scheme of our polarimeter was described above but to fully understand the measurements we are going to present, we must here include some extra details.
\subsection{Mirror birefringence}
\label{mirrorbirif}
The most important point is that the mirrors of Fabry-Perot cavities always present an intrinsic structured birefringence \cite{micossi} over the reflecting surface. The composition of the birefringence of the two mirrors can be treated as a single birefringent element \cite{brandi} inside a perfect non birefringent cavity. If $\alpha_1$ and $\alpha_2$ are the phase retardations upon reflection on each mirror for two polarisations parallel and perpendicular to their slow axis and if we take as a reference angle the slow axis of the first mirror then, \emph{per round trip}, the two mirrors are equivalent to a single birefringent element with a total retardation $\alpha_{\rm EQ}$ at an angle $\vartheta_{\rm EQ}$ with respect to the first mirror's slow axis \cite{brandi}:
\begin{eqnarray}
\alpha_{\rm EQ} = \sqrt{(\alpha_1-\alpha_2)^2+4\alpha_1\alpha_2\cos^2\vartheta_{\rm WP}}\\
\cos2\vartheta_{\rm EQ} = \frac{\alpha_1+\alpha_2\cos2\vartheta_{\rm WP}}{\sqrt{(\alpha_1-\alpha_2)^2+4\alpha_1\alpha_2\cos^2\vartheta_{\rm WP}}}.
\end{eqnarray}
Typical values for $\alpha_{1,2}\sim 10^{-7} - 10^{-6}$ combined with $N\approx 4\times 10^5$ result in a total retardation of the cavity of
\[
\alpha_{\rm cav} = \frac{N\alpha_{\rm EQ}}{2}\sim 10^{-1}.
\]
This leads to a high finesse cavity having two non degenerate resonances slightly separated in frequency by 
\[
\Delta\nu_{\rm sep} = \nu_{fsr}\frac{\alpha_{\rm EQ}}{2\pi}
\]
where $\nu_{fsr} = \frac{c}{2D}$ is the cavity's free spectral range. This separation is to be compared with each resonance's FWHM $\Delta\nu_{\rm cav} = \nu_{fsr}/{\cal F}$.

To reach a good extinction, necessary to have a good sensitivity, the input polarisation must be aligned to one of the axes of the cavity's equivalent birefringence. In this way no component perpendicular to $E_\parallel$ will be generated by the cavity itself. With this alignment, the reflected light used to lock the laser to the cavity has therefore a polarisation parallel to the input polariser. For this reason the laser is locked to only one of the two resonances whereas the ellipticity (or rotation) signal will respond to the resonance shifted by $\Delta\nu_{\rm sep}$. 

As discussed in references \cite{Zavattini2006,DellaValle2015} the ratio $\Delta\nu_{\rm sep}/\Delta\nu_{\rm cav} = {\cal F}\frac{\alpha_{\rm EQ}}{2\pi}$ leads to an extra phase between the two perpendicular polarisation states and to a reduction of the signal. The resulting expression for $E_\perp(t)$ is therefore:
\[
E_\perp(t) = E_0\left[i\Psi_0  \left(1-i\frac{N\alpha_{\rm EQ}}{2}\right)k(\alpha_{\rm EQ})\sin2\vartheta(t) + i\eta(t)\right]
\]
(here we have assumed that there are no rotations $\Phi$) where 
\[
k(\alpha_{\rm EQ}) = \frac{1}{1+N^2\sin^2(\alpha_{\rm EQ}/2)}.
\]
The expression for $E_\perp(t)$ is actually only valid in the limit of low frequencies with $\nu\ll\Delta\nu_{\rm cav}$, as will be discussed in section \ref{time}.
Apart from a reduction of the ellipticity signal by a factor $k(\alpha_{\rm EQ})$, $E_\perp(t)$ is no longer a pure imaginary number but also has a real component. The result of this modification of $E_\perp(t)$ is a mixing of ellipticity and rotation by a factor $\frac{N\alpha_{\rm EQ}}{2}$.

In general in the presence of both a rotation $\Phi_0 = N\phi_0$ and an ellipticity $\Psi_0 = N\psi_0$ 
the measured ellipticity and the measured rotation will be
\begin{eqnarray}
\label{eq:mix}
\Psi^{\rm meas} &=& k(\alpha_{\rm EQ})\left[\Psi_0 -\frac{N\alpha_{\rm EQ}}{2}\Phi_0 \right]\nonumber\\
\Phi^{\rm meas} &=& k(\alpha_{\rm EQ})\left[\Phi_0 +\frac{N\alpha_{\rm EQ}}{2}\Psi_0 \right].
\end{eqnarray}  
Measuring $\alpha_{\rm EQ}$ is thus fundamental to disentangle ellipticities and rotations.

In the presence of a pure ellipticity $(\Phi_0 = 0)$ and having measured the finesse, one finds that
\begin{equation}
\left.\frac{\Phi^{\rm meas}}{\Psi^{\rm meas}}\right|_{\Phi_0 = 0} = \frac{N\alpha_{\rm EQ}}{2}
\end{equation}
gives a direct value for $\alpha_{\rm EQ}$. The same is true in the presence of a pure rotation $(\Psi_0 = 0)$ in which case 
\begin{equation}
\left.\frac{\Psi^{\rm meas}}{\Phi^{\rm meas}}\right|_{\Psi_0 = 0} = -\frac{N\alpha_{\rm EQ}}{2}.
\end{equation}
The determination of $\alpha_{\rm EQ}$ can therefore be easily done either by measuring a Cotton-Mouton signal or a Faraday effect with and without the quarter-wave plate inserted. 

As we will see below, the mixing of an ellipticity with a rotation will help us understand the origin of the excess noise typically observed in polarimeters based on high finesse cavities.

\subsection{Frequency response}
\label{time}
An ideal Fabry-Perot behaves as a first order low pass filter with a frequency cutoff $\nu_{\rm cut}$ determined by the cavity line width $\Delta\nu_{\rm cav}$:
\[
\nu_{\rm cut} = \frac{\Delta\nu_{\rm cav}}{2} = \frac{c}{4D{\cal F}}.
\]
Therefore in the presence of a non birefringent cavity the measurement of an ellipticity signal generated by a time dependent birefringence at a frequency $\nu$ will be filtered according to 
\[
h(\nu) = \frac{1}{\sqrt{1+\left(\frac{\nu}{\nu_{\rm cut}}\right)^2}} = \frac{1}{\sqrt{1+\left(\frac{2\pi\nu D N}{c}\right)^2}}.
\]
With a finesse ${\cal F} = 7\times 10^5$ and a Fabry-Perot length $D = 3.303\;$m, as is the case in the PVLAS experiment, the frequency cutoff is $\nu_{\rm cut} = 32\;$Hz. 

It can be shown \cite{apb2018} that for $N\alpha_{\rm EQ}/2\ll1$ the frequency response of the measured \emph{rotation} signal in the presence of a time dependent \emph{birefringence} (or vice-versa the ellipticity signal in the presence of an effect generating a pure rotation) is well approximated by
\[
H(\nu) = h(\nu)^2 = \frac{1}{{1+\left(\frac{\nu}{\nu_{\rm cut}}\right)^2}}.
\]
The expressions given in equations (\ref{eq:mix}) therefore become
\begin{eqnarray}
\label{eq:mix1}
\Psi^{\rm meas} &=& k(\alpha_{\rm EQ})h(\nu)\left[\Psi_0 -\frac{N\alpha_{\rm EQ}}{2}\Phi_0 h(\nu)\right]\nonumber\\
\Phi^{\rm meas} &=& k(\alpha_{\rm EQ})h(\nu)\left[\Phi_0 +\frac{N\alpha_{\rm EQ}}{2}\Psi_0 h(\nu)\right].
\end{eqnarray}
Significant filtering is therefore present already for frequencies $\nu \lesssim \nu_{\rm cut}$.

If $N\alpha_{\rm EQ}\lesssim 1$, as is the case under consideration, the first and second order filters of the Fabry-Perot deviate significantly from the standard curves \cite{apb2018}. Remembering that $\Psi_0= N\psi_0$ and that similarly $\Phi_0 = N\phi_0$, the products $Nk(\alpha_{\rm EQ})h(\nu)$ and $Nk(\alpha_{\rm EQ})h^2(\nu)\frac{N\alpha_{\rm EQ}}{2}$must be substituted with more complicated expressions:

\begin{small}
\begin{eqnarray}
&&Nk(\alpha_{\rm EQ})h(\nu) \quad\rightarrow \quad Nk(\alpha_{\rm EQ})h_{\alpha_{\rm EQ}}(\nu)  = \nonumber\\
\label{eq:factor}
&&= \sqrt{\frac{4\,[1-R\cos\alpha_{\rm EQ}(2\cos\delta-R\cos\alpha_{\rm EQ})]}{\left[1+R^2-2R\cos(\alpha_{\rm EQ}-\delta)\right]\left[1+R^2-2R\cos(\alpha_{\rm EQ}+\delta)\right]}} \\\nonumber \\
&&Nk(\alpha_{\rm EQ})h(\nu)^2\frac{N\alpha_{\rm EQ}}{2} \quad\rightarrow \quad 
\nonumber\\
&&\rightarrow \sqrt{\frac{4\,R^2\sin^2\alpha_{\rm EQ}}{\left[1+R^2-2R\cos(\alpha_{\rm EQ}-\label{eq:factor1}
\delta)\right]\left[1+R^2-2R\cos(\alpha_{\rm EQ}+\delta)\right]}}
\end{eqnarray}
\end{small}
where $\delta = \frac{2\pi\nu}{\nu_{fsr}}$ and $R$ is the reflectance of the mirrors (assumed to be equal). Furthermore it can be shown that, to order $\alpha_{\rm EQ}^2$, the ratio of equation (\ref{eq:factor1}) to equation (\ref{eq:factor}) is
\begin{equation}
\sqrt{\frac{R^2\sin^2\alpha_{\rm EQ}}{1-R\cos\alpha_{\rm EQ}(2\cos\delta-R\cos\alpha_{\rm EQ})}} = \frac{N\alpha_{\rm EQ}}{2}h(\nu)
\end{equation}
and is therefore proportional to a simple first order filter independent of $\alpha_{\rm EQ}$.

With all these considerations the measured values for $\Psi^{\rm meas}$ and $\Phi^{\rm meas}$ are 
\begin{eqnarray}
\label{eq:mix2}
\Psi^{\rm meas} &=& k(\alpha_{\rm EQ})h_{\alpha_{\rm EQ}}(\nu)\left[\Psi_0 -\frac{N\alpha_{\rm EQ}}{2}\Phi_0 h(\nu)\right]\nonumber\\
\Phi^{\rm meas} &=& k(\alpha_{\rm EQ})h_{\alpha_{\rm EQ}}(\nu)\left[\Phi_0 +\frac{N\alpha_{\rm EQ}}{2}\Psi_0 h(\nu)\right].
\end{eqnarray}  
Therefore in the presence of a pure birefringence $(\Phi_0 = 0)$ and having measured the finesse, one finds that
\begin{equation}
\label{eq:ratio1}
\left.\frac{\Phi^{\rm meas}}{\Psi^{\rm meas}}\right|_{\Phi_0 = 0} = \frac{N\alpha_{\rm EQ}}{2}{h(\nu)}
\end{equation}
gives a direct value for $\alpha_{\rm EQ}$. 
The same is true in the presence of an effect generating a pure rotation $(\Psi_0 = 0)$ in which case 
\begin{equation}
\label{eq:ratio2}
\left.\frac{\Psi^{\rm meas}}{\Phi^{\rm meas}}\right|_{\Psi_0 = 0} = -\frac{N\alpha_{\rm EQ}}{2}h(\nu).
\end{equation}
\subsection{Noise studies}
Since the measured noise both in ellipticity and in rotation is significantly greater than the expected noise, we assume independent contributions by both ellipticity and/or rotation noises generated and amplified inside the cavity.
We model the measured spectral noise densities as

\begin{small}
\begin{eqnarray}
\label{eq:mixnoise}
&& S_{\Psi^{\rm meas}}(\nu)= k(\alpha_{\rm EQ})h_{\alpha_{\rm EQ}}(\nu)\times\\
&&\times\sqrt{S_\Psi(\nu)^2 +\left(\frac{N\alpha_{\rm EQ}}{2}S_\Phi(\nu) h(\nu)\right)^2+\left(\frac{S_e}{k(\alpha_{\rm EQ})h_{\alpha_{\rm EQ}}(\nu)}\right)^2}\nonumber\\\nonumber\\
&&S_{\Phi^{\rm meas}}(\nu)=k(\alpha_{\rm EQ})h_{\alpha_{\rm EQ}}(\nu)\times \\
&&\times\sqrt{S_\Phi(\nu)^2 +\left(\frac{N\alpha_{\rm EQ}}{2}S_\Psi(\nu) h(\nu)\right)^2+\left(\frac{S_r}{k(\alpha_{\rm EQ})h_{\alpha_{\rm EQ}}(\nu)}\right)^2}\nonumber
\end{eqnarray}  
\end{small}
where $S_\Psi(\nu)=Ns_\psi(\nu)$ and $S_\Phi(\nu)=Ns_\phi(\nu)$ are respectively the ellipticity and rotation spectral densities and where we have added white noise contributions $S_r$ and $S_e$ to $S_{\Phi^{\rm meas}}(\nu)$ and $S_{\Psi^{\rm meas}}(\nu)$ respectively. 

By studying the noise spectra in ellipticity and rotation and the Cotton-Mouton and Faraday signals one can confirm whether the noises $S_\Psi(\nu)$ and $S_\Phi(\nu)$ are proportional to $N$ or not and therefore if they originate from inside or outside of the cavity. Finally by comparing the measured ellipticity noise $S_{\Psi^{\rm meas}}(\nu)$ with the measured rotation noise $S_{\Phi^{\rm meas}}(\nu)$ one can determine whether these are dominated by $S_\Psi(\nu)$ or $S_\Phi(\nu)$.

\section{Measurements}
Let us remind the reader that the aim of the present work is to study the signal-to-noise ratio in the PVLAS apparatus as a function of the finesse of the Fabry-Perot cavity. To reduce the finesse of the cavity we have introduced controlled extra losses $p$ to the Fabry-Perot cavity. Given the transmittance $T$ and the intrinsic losses $p_0$ of the mirrors, $p$ will cause the finesse ${\cal F}$ and the output intensity $I_0$ to change according to
\begin{eqnarray}
{\cal F}(p) &=& \frac{\pi}{T+p_0+p}\\
\frac{I_0(p)}{I_{\rm in}} &=& \left[\frac{T}{T+p_0+p}\right]^2 = \left[\frac{T{\cal F}(p)}{\pi}\right]^2.
\end{eqnarray}

In the case of the PVLAS cavity, the best finesse measured was ${\cal F} \approx 7.7\times 10^5$ with a 25\% transmission \cite{OE2014}, corresponding to $p_0 = (1.7\pm0.2)\;$ppm and a transmittance of each mirror (assumed to be equal) $T = 2.4\pm0.2\;$ppm. Therefore an extra loss $p\approx0\div10$~ppm will change the finesse from ${\cal F} = 7.7\times 10^5$ to ${\cal F} =2.5 \times 10^5$. 

To introduce these extra losses we have used one of the manual vacuum gate valves present in front of the output mirror to clip the Gaussian mode between the mirrors. With a width $r_0$ of the intensity profile of the Gaussian mode and therefore $\sigma= \frac{r_0}{2}$, clipping at $(4.5-5) \sigma$ level is sufficient to achieve the desired losses $p$. An estimate can be made considering a circular aperture of radius $a$. The power loss per pass of the beam inside the cavity is
\[
p \approx e^{-2\frac{a^2}{w^2}} = e^{-\frac{a^2}{2\sigma^2}}.
\]
With a ratio $x = a/\sigma = 4.8$ the resulting extra losses are $p = 10\;$ppm. Given the relatively large value of $x$, these extra power losses are therefore obtained without significantly altering the Gaussian beam profile. It is also true, though, that very small position variations of the gate valve with respect to the beam generate significant variations of the finesse. We have observed that, with the valve inserted, the stability of the finesse is of the order of 1\%. In our measurements this is the dominant uncertainty factor.

Note also that even if the noise is generated within the whole thickness of the reflecting layers, the physical structures of the multilayer dielectric mirrors corresponding to the highest and the lowest finesses used would differ by no more than a pair of dielectric layers \cite{Born}; this justifies the use of an extra loss located outside the mirrors as a means to study the intrinsic birefringence noise of the mirrors as a function of the finesse.

Since a reduction by a factor 3 of the finesse results in a factor 9 reduction at the output of the cavity and given that the output power is $I_{\rm cav} = T I_\parallel$ this also means a factor 9 reduction of the power on the mirrors. We therefore chose to change the input power to the cavity so that at each finesse the output power was the same during all measurements: we chose $I_0 = 0.7\;$mW.

The theoretical sensitivity for $I_0 = 0.7\;$mW was already shown in figure \ref{fig:sens}. Superimposed is also the measured sensitivity between 10~Hz and 20~Hz with a modulation $\eta_0 = 10^{-2}$. This measured sensitivity does not change by increasing or decreasing the output power by a factor ten.

%
During our measurements the magnets were kept in rotation at two different frequencies, $\nu_\alpha = 4\;$Hz and $\nu_\beta=5\;$Hz generating Cotton-Mouton peaks at twice these frequencies due to the presence of Argon gas at $850\;\mu$bar. The frequency of the Faraday rotation signal induced on the input mirror using an external solenoid was chosen to be $\nu_F = 19\;$Hz.

Six different values of the finesse were chosen for the measurements, each separated by approximately 20\%. 
For each finesse value we first measured in the ellipticity configuration and then in the rotation configuration by inserting the quarter-wave plate. The finesse was determined by measuring the intensity decay exiting the cavity after unlocking the laser at the end of each series of measurements.

\begin{figure}[tb]
\begin{center}
\includegraphics[width=8.5cm]{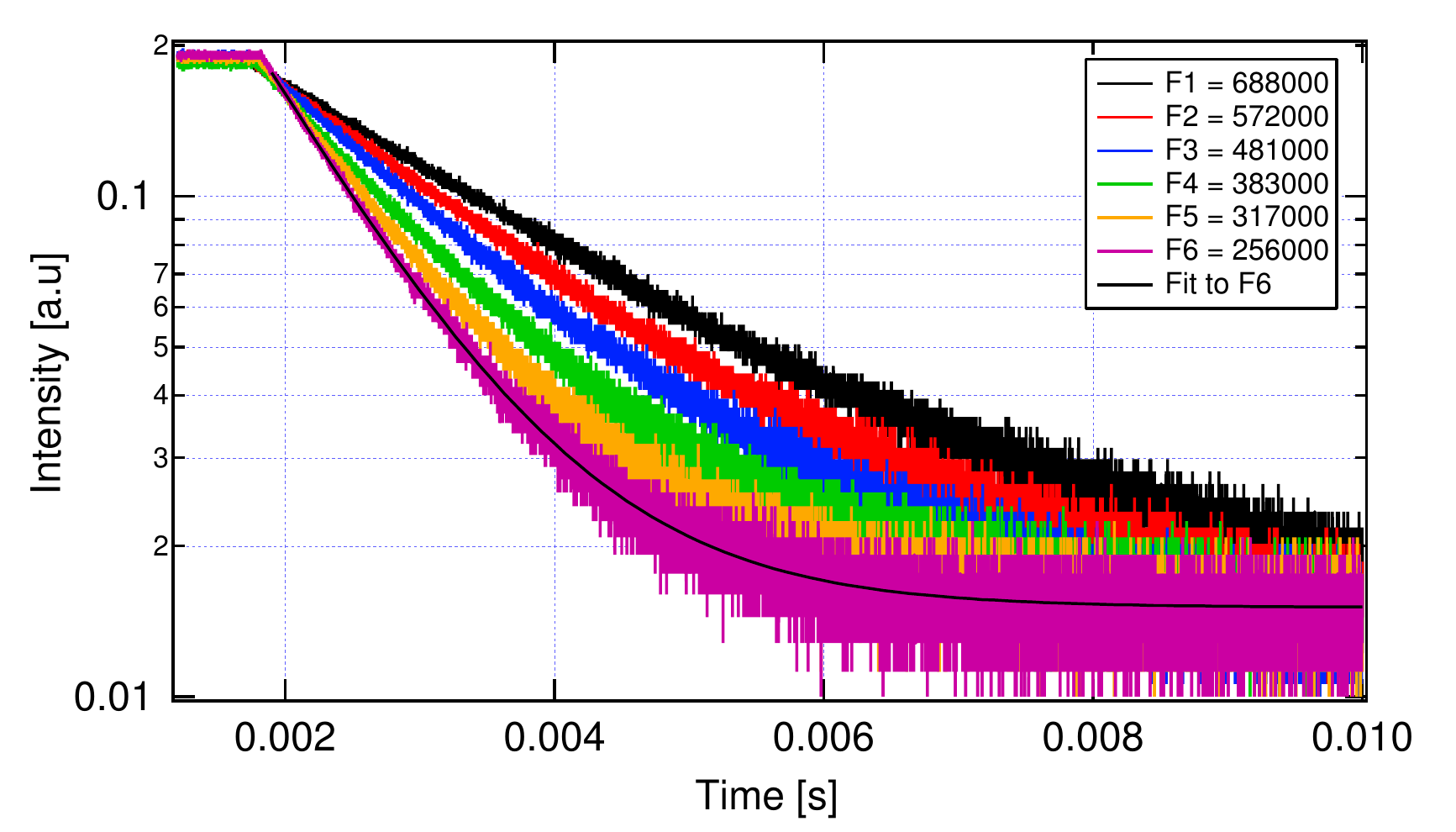}
\end{center}
\caption{Light decay curves for the six different positions of the gate valve clipping the beam to increases the losses inside the cavity. F1 - F6 represent the relative finesse values. As an example, an exponential fit is superimposed to the curve relative to F6.}
\label{fig:decay}
\end{figure}

The intensity decay graphs for the six positions of the gate valve, resulting in the six finesse values used during the measurements, are shown in figure \ref{fig:decay}.
These correspond respectively to 
\[
{\cal F} = (6.88, 5.72, 4.81, 3.83, 3.17, 2.56)\times 10^5
\]
with a 1\% uncertainty.

The main goal of the present work is to show whether the noise present in the two configurations of the polarimeter is dominated by an ellipticity noise generated by a fluctuating birefringence inside the cavity, i.e. whether it is multiplied by the gain factor $N$ of the Fabry-Perot. To accomplish this, for each finesse value we first determined the value of $\alpha_{\rm EQ}$ from equations (\ref{eq:ratio1}) and (\ref{eq:ratio2}) for both the Faraday and the Cotton-Mouton measurements. The dependence of both the Cotton-Mouton and Faraday signals are expected to follow the relations given in equations (\ref{eq:mix2}). We have then studied the signal-to-noise ratios of the various signals both in the rotation and ellipticity configurations to study their behaviour as a function of the finesse.


\section{Results and discussion}
\begin{figure}[b]
\begin{center}
\includegraphics[width=8.5cm]{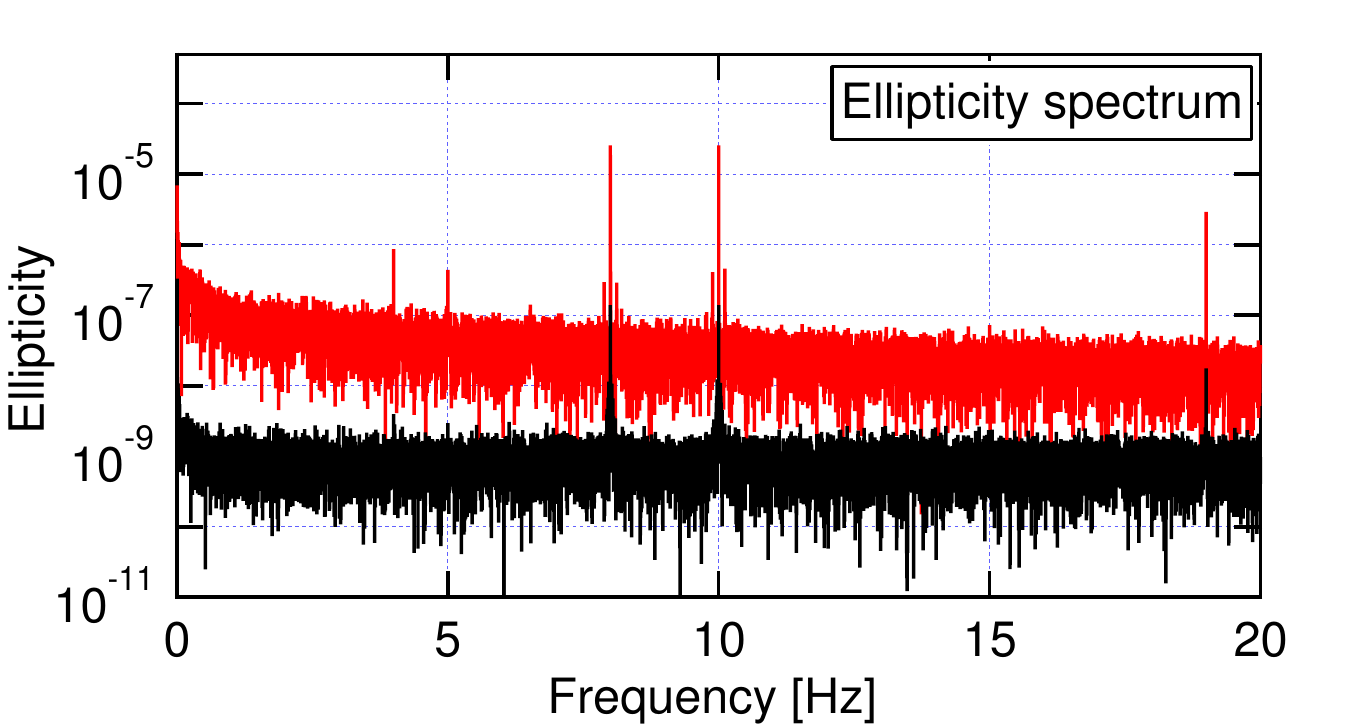}
\includegraphics[width=8.5cm]{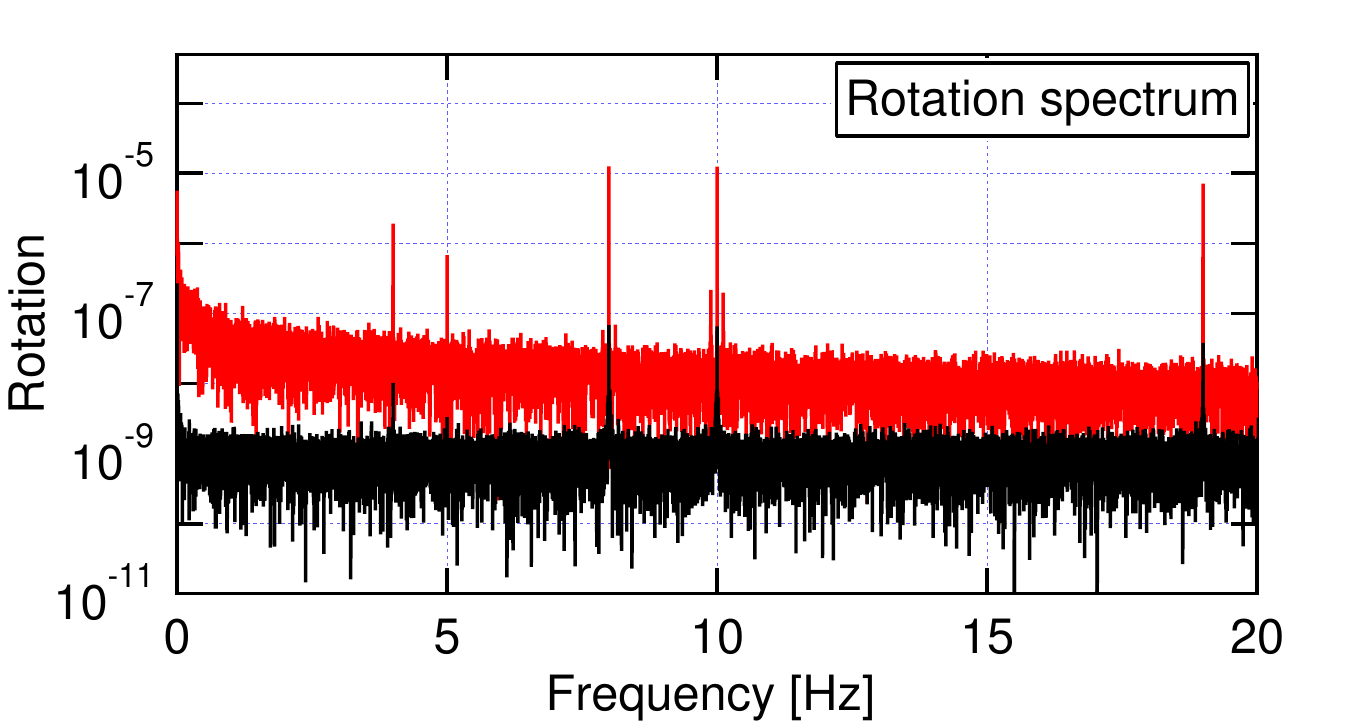}
\end{center}
\caption{Ellipticity (top panel) and rotation (bottom panel) raw spectra for an integration time of $t = 512\;$s for ${\cal F}_1 = 6.88\times 10^5$. In red is the FFT of the in-phase component whereas in black is the quadrature component. A zoom from 0~Hz to 20~Hz is shown to better appreciate the peaks at $2\nu_\alpha = 8\;$Hz, $2\nu_\beta = 10\;$Hz, of equal amplitudes, and at $\nu_F = 19\;$Hz. Peaks at $\nu_\alpha$ and $\nu_\beta$ are also present due to a slight non orthogonality between the beam and the magnetic field direction generating a time dependent Faraday rotation in the Argon gas. 
}
\label{fig:spettro}
\end{figure}

Typical raw ellipticity (top) and rotation (bottom) spectra measured in a time $t = 512\;$s at the highest finesse  of ${\cal F}_1 = 6.88\times 10^5$ are shown in figure \ref{fig:spettro}. Data sampling was performed at 256 Hz resulting in spectra with a frequency resolution of $\Delta\nu_{\rm res} = 1.5\times 10^{-5}\;$Hz. In both the ellipticity and rotation channels the Argon Cotton-Mouton signals at $2\nu_{\alpha} = 8\;$Hz and $2\nu_{\beta} = 10\;$Hz are clearly visible, with equal amplitudes due to identical magnets, along with the Faraday rotation signal at $\nu_F = 19\;$Hz induced in the input mirror of the cavity. The small sidebands around $2\nu_{\alpha,\beta}$ are due to small oscillations of the rotation frequency of the magnets generated by the driving system of the magnets. The amplitude error on the main peaks due to these sidebands is less than $1\;$\textperthousand. 
In both panels of the figure \ref{fig:spettro} one can also distinguish two peaks at $\nu_{\alpha} = 4\;$Hz and $\nu_{\beta} = 5\;$Hz due to a small component of the magnetic field along the beam direction generated by a small non orthogonality of the magnetic field with respect to the beam propagation direction. This small component of the magnetic field generates a Faraday rotation in the gas inside the cavity. Indeed these peaks are higher in the rotation spectrum. A small Faraday effect is also generated in the mirrors due to the stray field but this rotation is negligible with respect to the rotation generated in the gas.

Notice how the integrated noise and the two Cotton-Mouton signals are smaller in the rotation spectrum with respect to the ellipticity spectra whereas the Faraday signals are larger in the rotation spectrum. 

In figure \ref{fig:spettro} we have also reported the quadrature demodulation spectra integrated over the same time $t$. This integrated noise corresponds to a peak spectral density of $S_{\rm quad} = 1.6\times 10^{-8}\;1/\sqrt{\rm Hz}$, in agreement with the sensitivity, shown in figure \ref{fig:sens}, due to noise sources independent of ellipticity such as shot-noise, Johnson noise, diode dark current noise and laser relative intensity noise considering $I_0 = 0.7\;$mW and $\eta_0 = 10^{-2}$. This noise is the same in both the ellipticity and the rotation spectrum. The in-phase noise is clearly of a different origin. Small peaks, less than 1\% of the in-phase peaks, are present in the quadrature channel (in black in figure \ref{fig:spettro}) due to a phase error in the demodulation of about $1^\circ$.
\begin{figure}[t]
\begin{center}
\includegraphics[width=8.5cm]{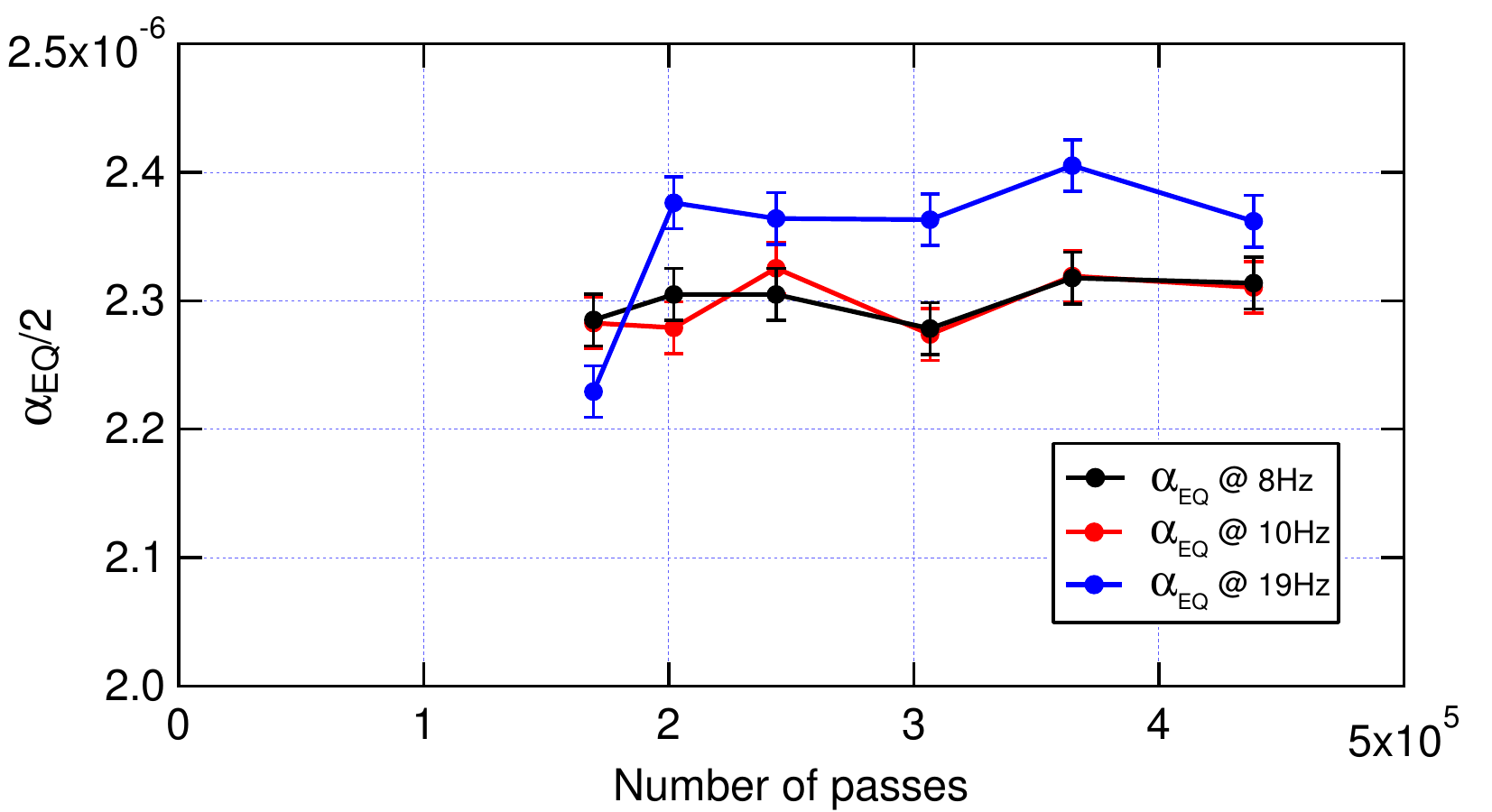}
\end{center}
\caption{Determination of $\alpha_{\rm EQ}$ as a function of $N$ for both the Cotton-Mouton signals at $2\nu_{\alpha,\beta}$ and for the Faraday rotation at $\nu_F$. 
}
\label{fig:AlphaEll}
\end{figure}


\subsection{Determination of $\alpha_{EQ}$}
For each value of the finesse we have first determined the value of $\alpha_{EQ}$, necessary to evaluate the true ellipticities and rotations due to the Cotton-Mouton and Faraday effects, according to equations (\ref{eq:ratio1}) and (\ref{eq:ratio2}). In figure \ref{fig:AlphaEll} we have plotted the values of the ratios
\[
\left.\frac{\Phi^{\rm meas}}{\Psi^{\rm meas}}\frac{2}{Nh(\nu)}\right|_{\nu = 8\;{\rm Hz},10\;{\rm Hz}} = \alpha_{\rm EQ}
\]
at 8~Hz and 10~Hz as a function of the number of passes $N = \frac{2{\cal F}}{\pi}$ and
\[
\left.\frac{\Psi^{\rm meas}}{\Phi^{\rm meas}}\frac{2}{Nh(\nu)}\right|_{\nu = 19\;{\rm Hz}} = \alpha_{\rm EQ}.
\]
Since $\alpha_{\rm EQ}$ is a property of the mirror coatings it is independent of $N$, as expected.
The average value for $\alpha_{\rm EQ}$ at 8~Hz and 10~Hz is
\[
\alpha_{\rm EQ}|_{\nu = 8\;{\rm Hz},10\;{\rm Hz}} = 2.30\times 10^{-6}\quad\quad\sigma_{\alpha_{\rm EQ}} = 2\times 10^{-8}
\]
where $\sigma_{\alpha_{\rm EQ}}$ is the standard deviation of each single measurement.
The same can be done by considering the measured rotation and ellipticity peaks at $\nu = 19\;$Hz:
\[
\alpha_{\rm EQ}|_{\nu = 19\;{\rm Hz}} = 2.35\times 10^{-6}\quad\quad\sigma_{\alpha_{\rm EQ}} = 6\times 10^{-8}.
\]

The two values are compatible within the experimental uncertainty. The slight tendency to increase with $N$ of the value obtained with the Faraday effect might be due to the small contribution of the mirror substrate. The weighted average of the two values, that we will use in the following, is $\alpha_{\rm EQ}=2.305\pm0.019$.
\begin{figure}[b]
\begin{center}
\includegraphics[width=7.5cm]{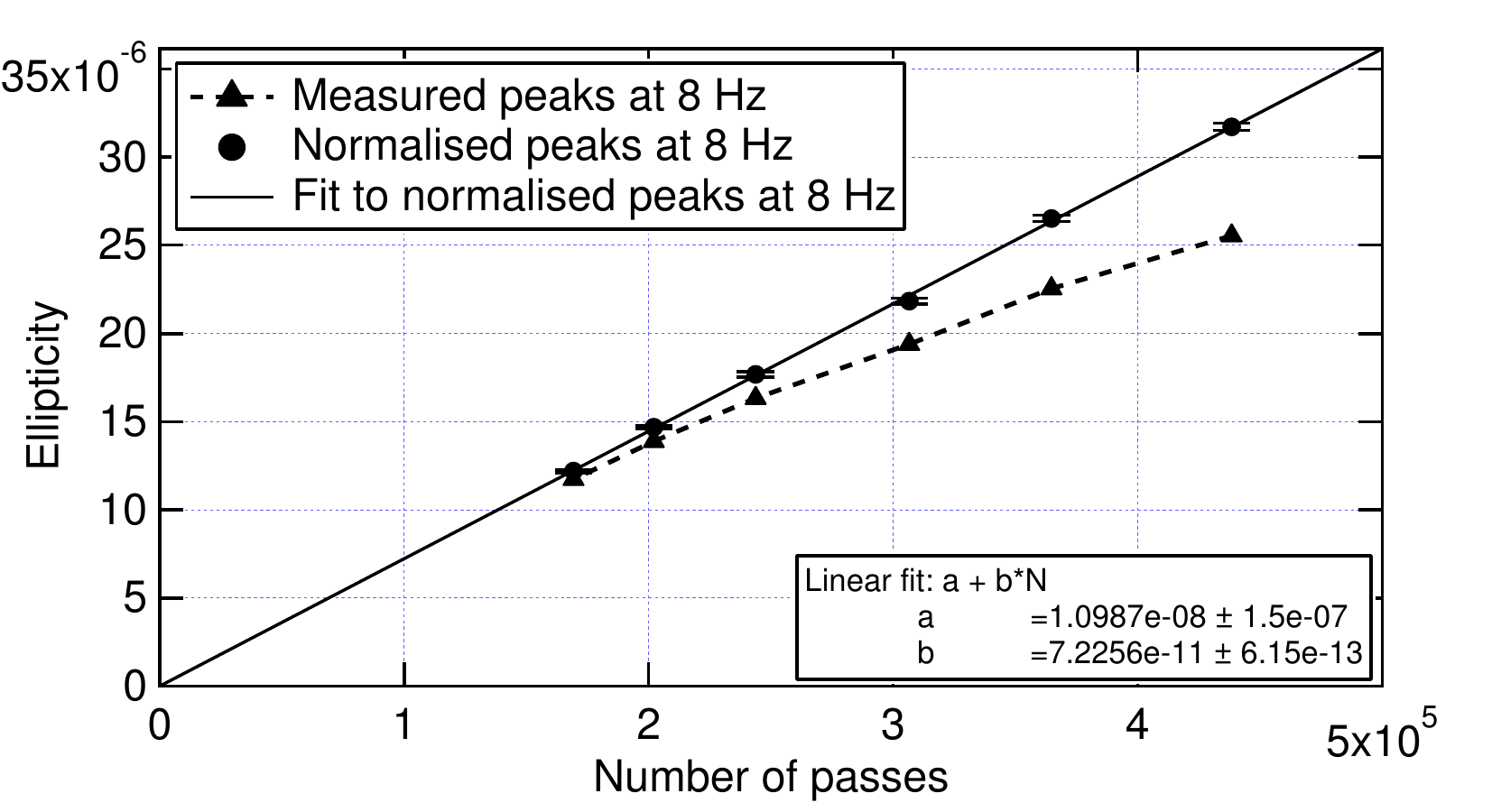}
\includegraphics[width=7.5cm]{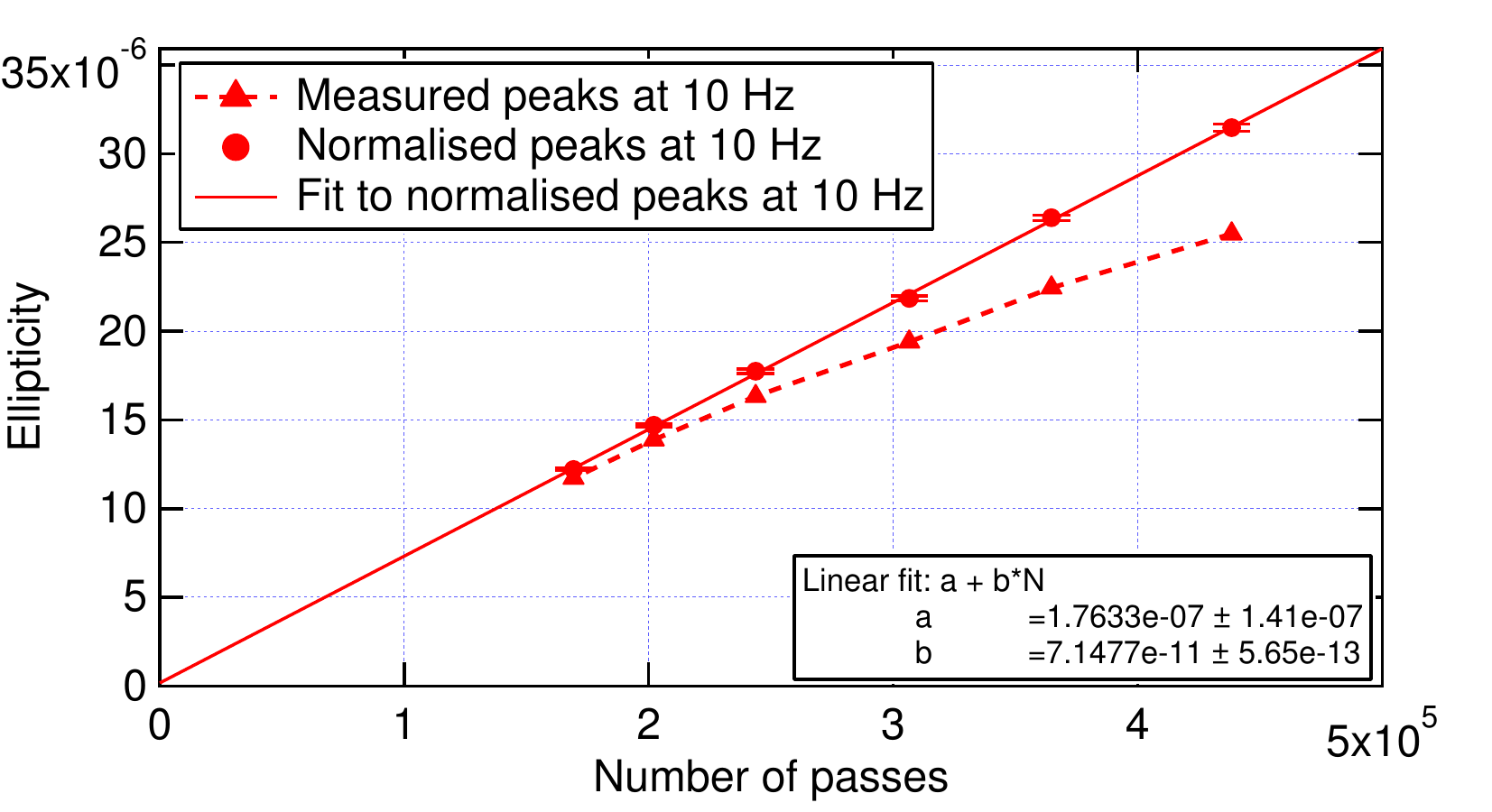}
\end{center}
\caption{Measured Cotton-Mouton peaks at $2\nu_{\alpha}$ (top) and $2\nu_{\beta}$ (bottom) as a function of $N$. The experimental points connected with dashed lines are not corrected for the cavity response. The dots lying on the linear fit are the experimental values corrected for the cavity response. The two slopes are compatible within their errors.}
\label{fig:CMvsN}
\end{figure}


\subsection{Cotton-Mouton and Faraday signals in the ellipticity channel versus $N$}

In figure \ref{fig:CMvsN} we have plotted the peak amplitudes $\Psi^{\rm meas}$ of the ellipticity signals at 8~Hz and 10~Hz as a function of $N$. In the same figure we have also plotted the values of $\Psi_0$ obtained from equation (\ref{eq:mix2}) taking into account the frequency dependence $h_{\alpha_{\rm EQ}}(\nu)$ of the signals and the amplitude reduction due to $k(\alpha_{\rm EQ})$. 
As expected these lie on a line passing through the origin indicating that indeed the signal is $\Psi_0 = N\psi_0$. The slope of the lines give the value of the ellipticity per pass $\psi_0 = (7.20\pm0.04)\times 10^{-11}$ acquired by the light resulting in a Cotton-Mouton constant \cite{rizzo} $\Delta n_u = {\Delta n}/{B^2} = (5.63\pm0.14)\times 10^{-15}\;$T$^{-2}$ @ 1064 nm with $P_{\rm Ar} = (850\pm20)\;\mu$bar.

\begin{figure}[t]
\begin{center}
\includegraphics[width=7.5cm]{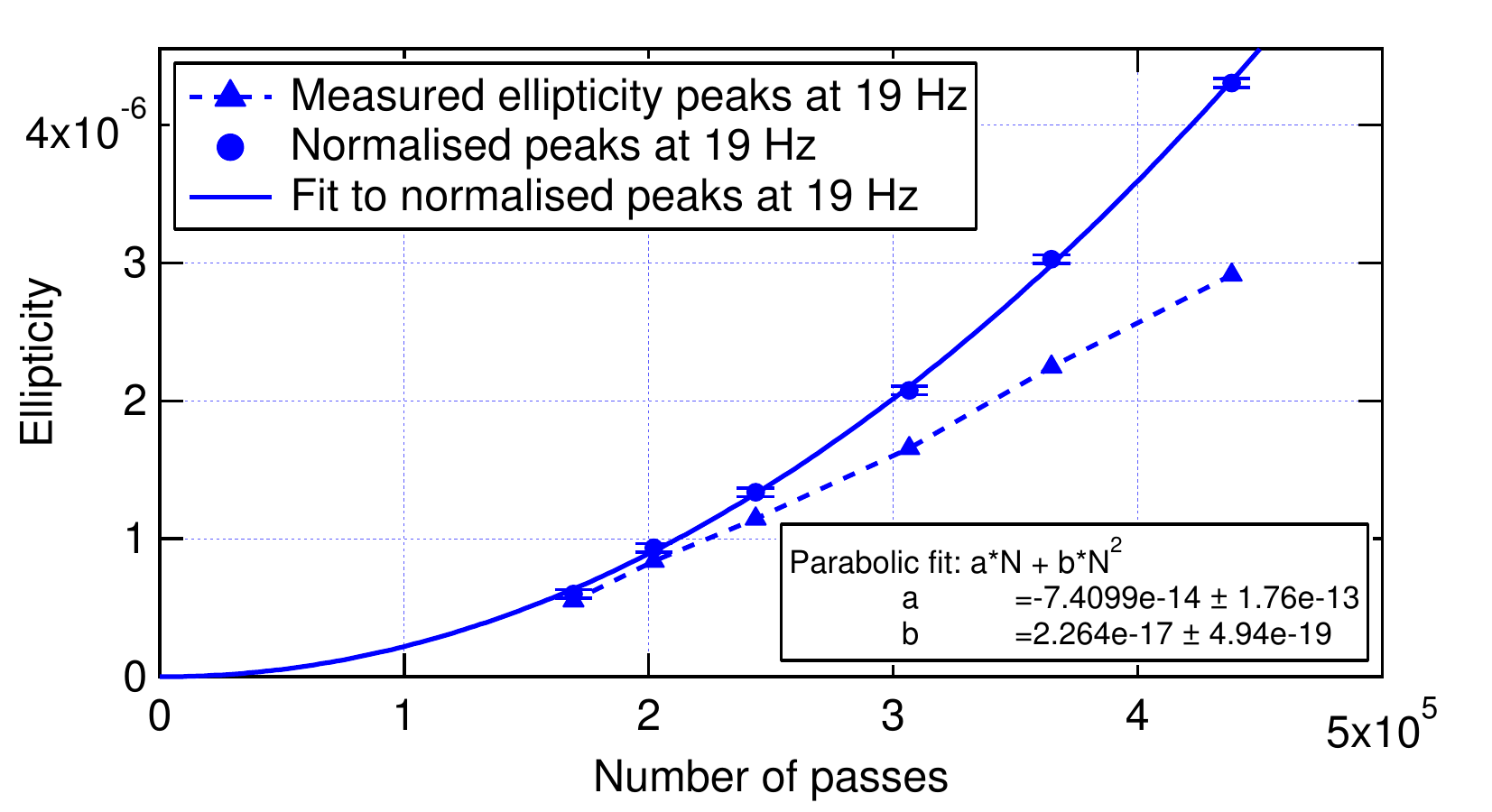}
\end{center}
\caption{Measured Faraday peak at $\nu_{F}$ in the ellipticity channel as a function of $N$. The experimental points connected with dashed lines are not corrected for the cavity response. The dots lying on the parabolic fit are the experimental values corrected for the cavity response.}
\label{fig:FvsNell}
\end{figure}
In figure \ref{fig:FvsNell} we have plotted the values of $\Psi^{\rm meas}$ of the ellipticity at 19~Hz. This signal is due to the Faraday rotation being transformed into ellipticity because of the birefringence of the cavity. In the same figure we have also plotted the values of $\frac{\alpha_{\rm EQ}}{2}\phi_0$, obtained having normalised the values $\Psi^{\rm meas}$ for the response of the polarimeter, as a function of $N$ according to the expression deduced from equation (\ref{eq:mix2}) having set $\Psi_0 = 0$:
\[
\left|\frac{\Psi^{\rm meas}}{h(\nu)h_{\alpha_{\rm EQ}}(\nu)k(\alpha_{\rm EQ})}\right|_{19\;{\rm Hz}}=\frac{N\alpha_{\rm EQ}}{2}\Phi_0 = N^2\frac{\alpha_{\rm EQ}}{2}\phi_0.
\]
As expected from this last equation these values lie on a parabolic curve allowing the determination of the rotation per pass $\phi_0 = (1.96\pm0.04)\times 10^{-11}\;$rad/pass. 
It is estimated that the contribution of the substrate is less than 1\%, compatible with the parabola passing through the origin within the errors.
\begin{figure}[t]
\begin{center}
\includegraphics[width=8.5cm]{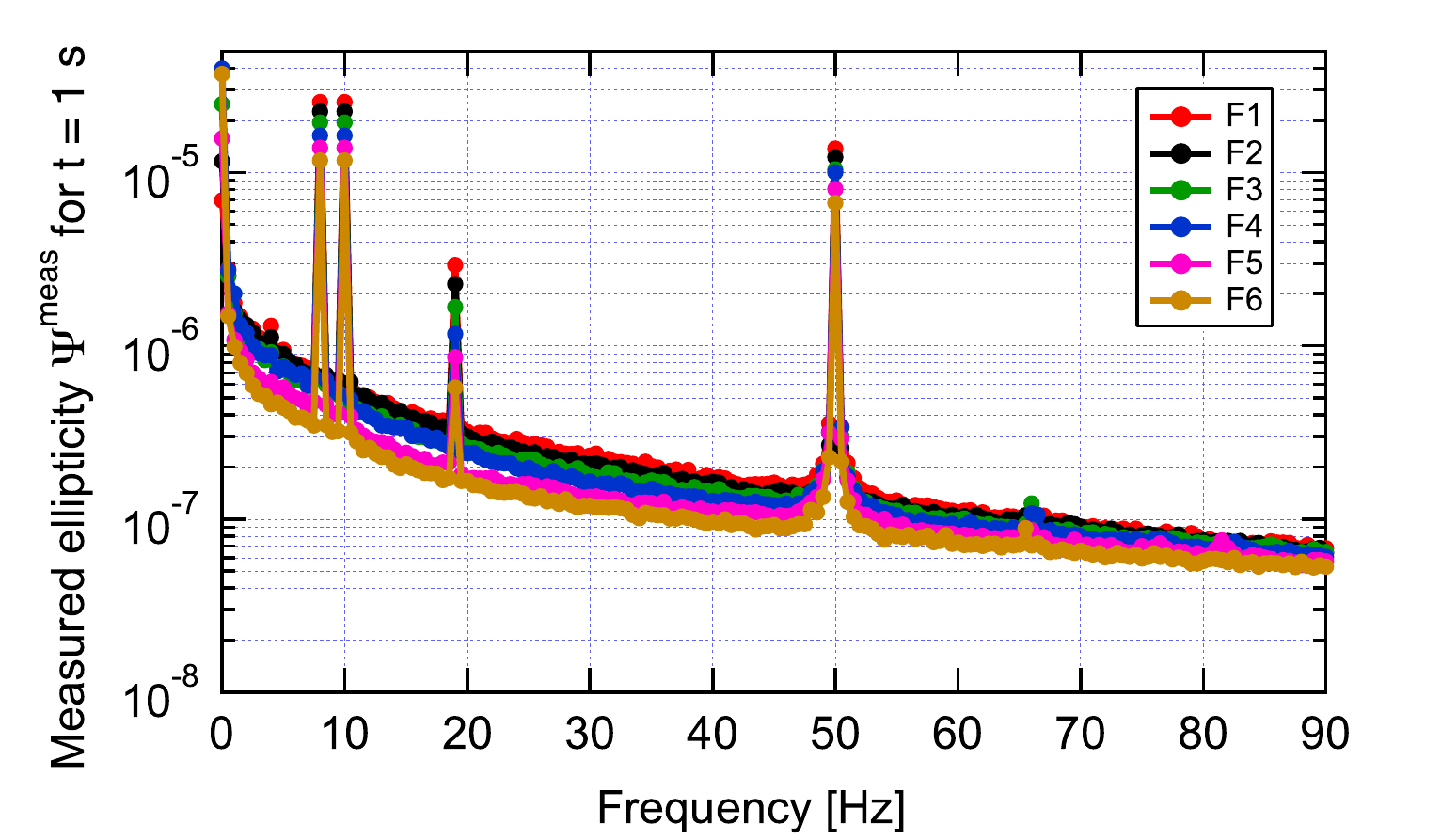}
\end{center}
\caption{Ellipticity spectra for the six finesse values rescaled to a 1~s integration time. The raw spectra have been rebinned by taking rms averages of the raw spectra in 0.5~Hz frequency intervals. The peak at 50 Hz is due to the mains.}
\label{fig:spettri_ell}
\end{figure}

\subsection{Ellipticity noise versus $N$}
In figure \ref{fig:spettri_ell} one can see the ellipticity rms spectra at the six different finesse values, where the raw spectra have been rebinned in 0.5~Hz frequency bins.

\begin{figure}[t]
\begin{center}
\includegraphics[width=8.5cm]{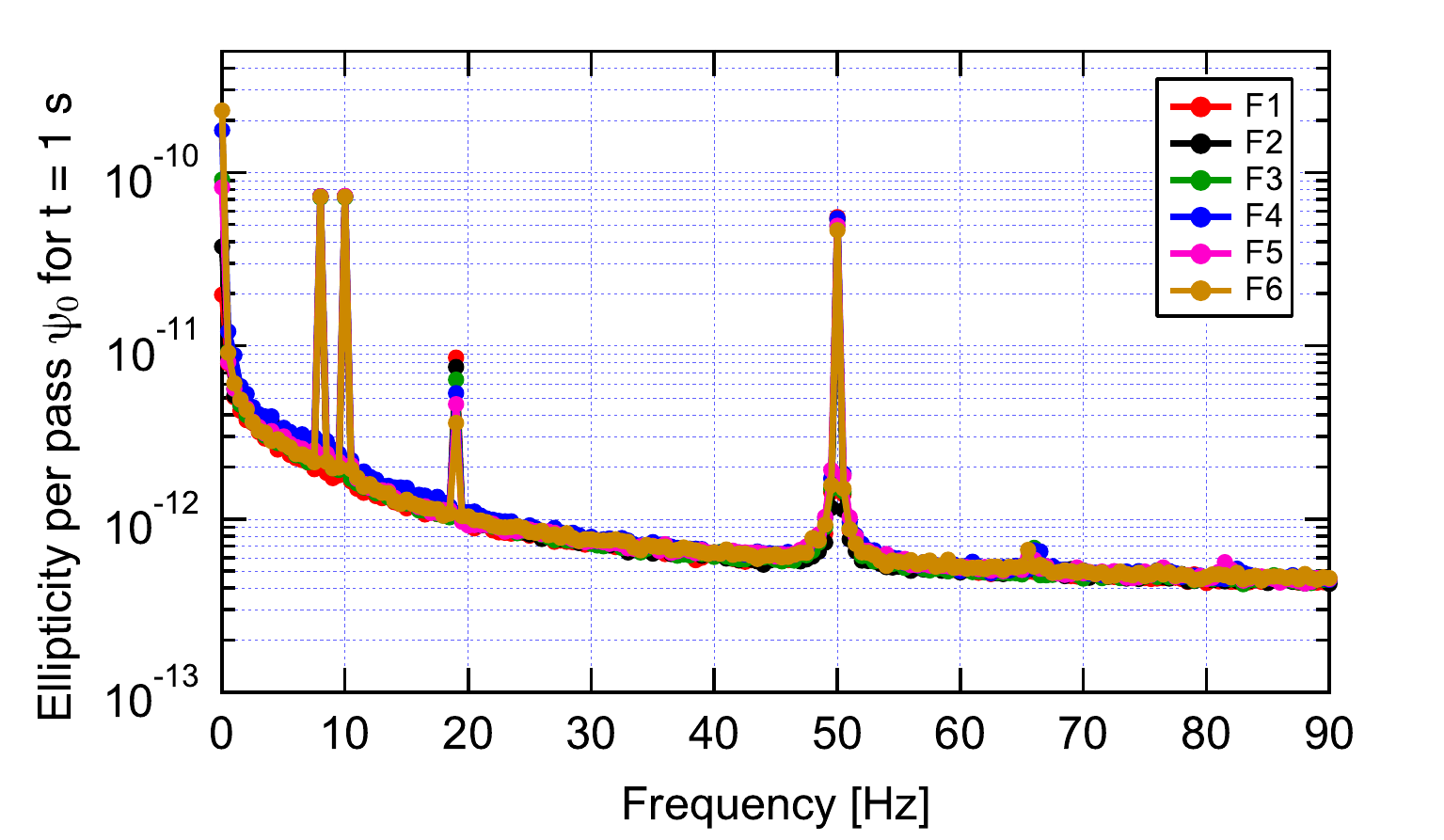}
\end{center}
\caption{The six ellipticity spectra of figure \ref{fig:spettri_ell} rescaled assuming an ellipticity noise $S_\Psi$ proportional to $N$ and taking into account the frequency response of the cavity.}
\label{fig:spettri_ell_norm}
\end{figure}

These spectra 
have not been renormalised for the amplification factor $N$ and frequency response $k(\alpha_{\rm EQ})h_{\alpha_{\rm EQ}}(\nu)$ given by equation (\ref{eq:factor}) due to the Fabry-Perot with equivalent birefringence per pass $\alpha_{\rm EQ}$. As can be seen not only do the Cotton-Mouton ellipticity signals and the peaks at $\nu_F = 19\;$Hz decrease with decreasing $N$, as already discussed, but so does the noise. By normalising each spectrum with the cavity response given by equation (\ref{eq:factor}) and assuming the noise to be dominated by the intracavity ellipticity noise $s_\psi$ ($s_\phi=0$ and $S_e=0$), one finds the plot shown in figure \ref{fig:spettri_ell_norm}.
\begin{figure}[t]
\begin{center}
\includegraphics[width=8.5cm]{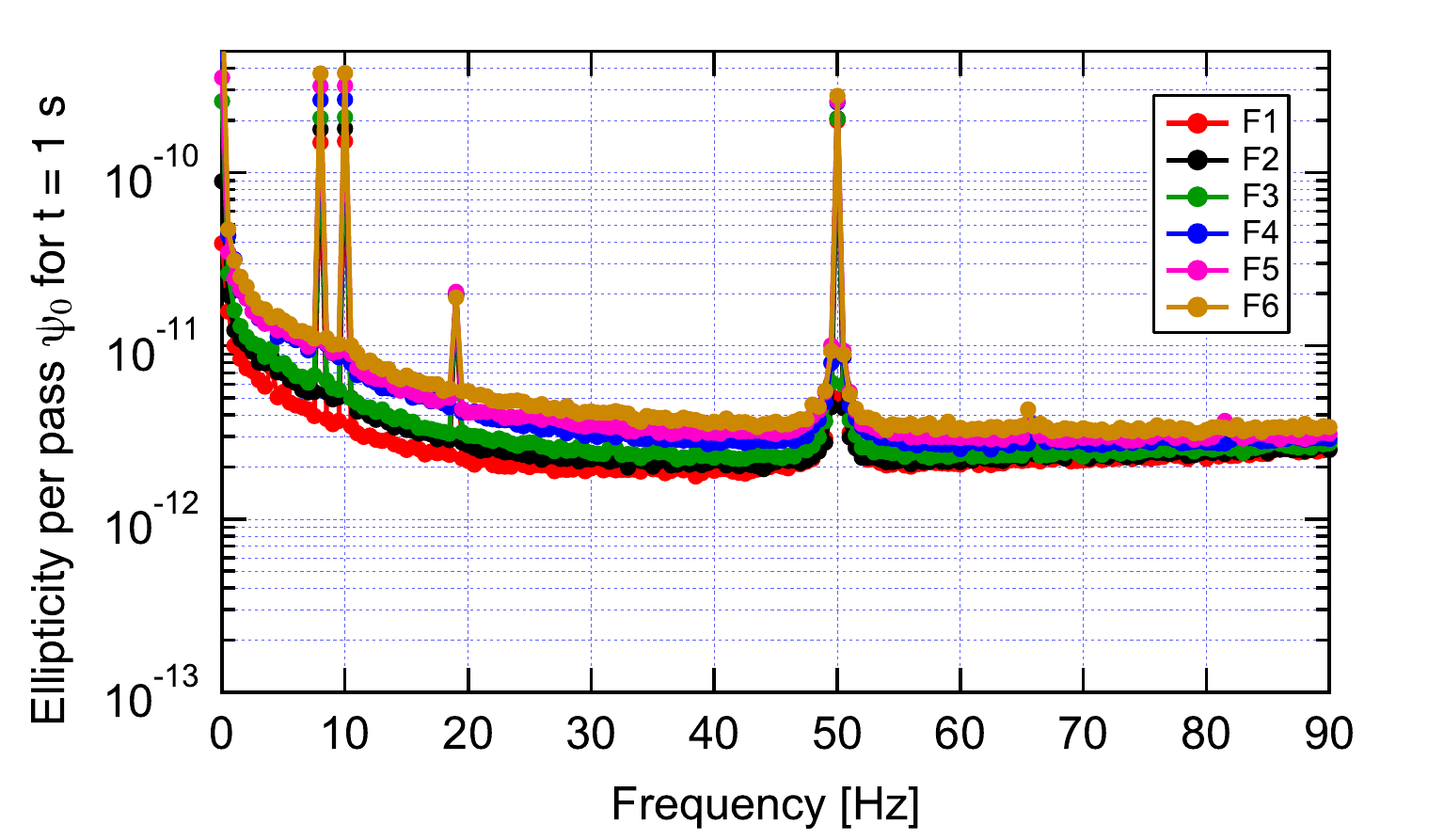}
\end{center}
\caption{The six ellipticity spectra of figure \ref{fig:spettri_ell} rescaled assuming a rotation noise $S_\Phi$ proportional to $N$ and taking into account the frequency response of the cavity.}
\label{fig:spettri_ell_norm_rot}
\end{figure}
In this figure, all the noise components of the spectra lie on a common curve (except for a small broad structure between 5~Hz and 10~Hz at the lower finesse values). The Cotton-Mouton peaks also indicate a common value whereas the signal at $\nu_F$ does not, as expected. 

Instead, by normalising the noise spectra assuming an intracavity rotation noise $s_\phi$ ($s_\psi=0$ and $S_e=0$), one obtains the plots in figure \ref{fig:spettri_ell_norm_rot}. In this case the peaks at $\nu_F$, which have an origin from a Faraday effect, overlap whereas the noise and the Cotton-Mouton peaks do not. It is also apparent that the noise does not behave as an intracavity rotation noise $s_\phi$.
\begin{figure}[htb]
\begin{center}
\includegraphics[width=8.5cm]{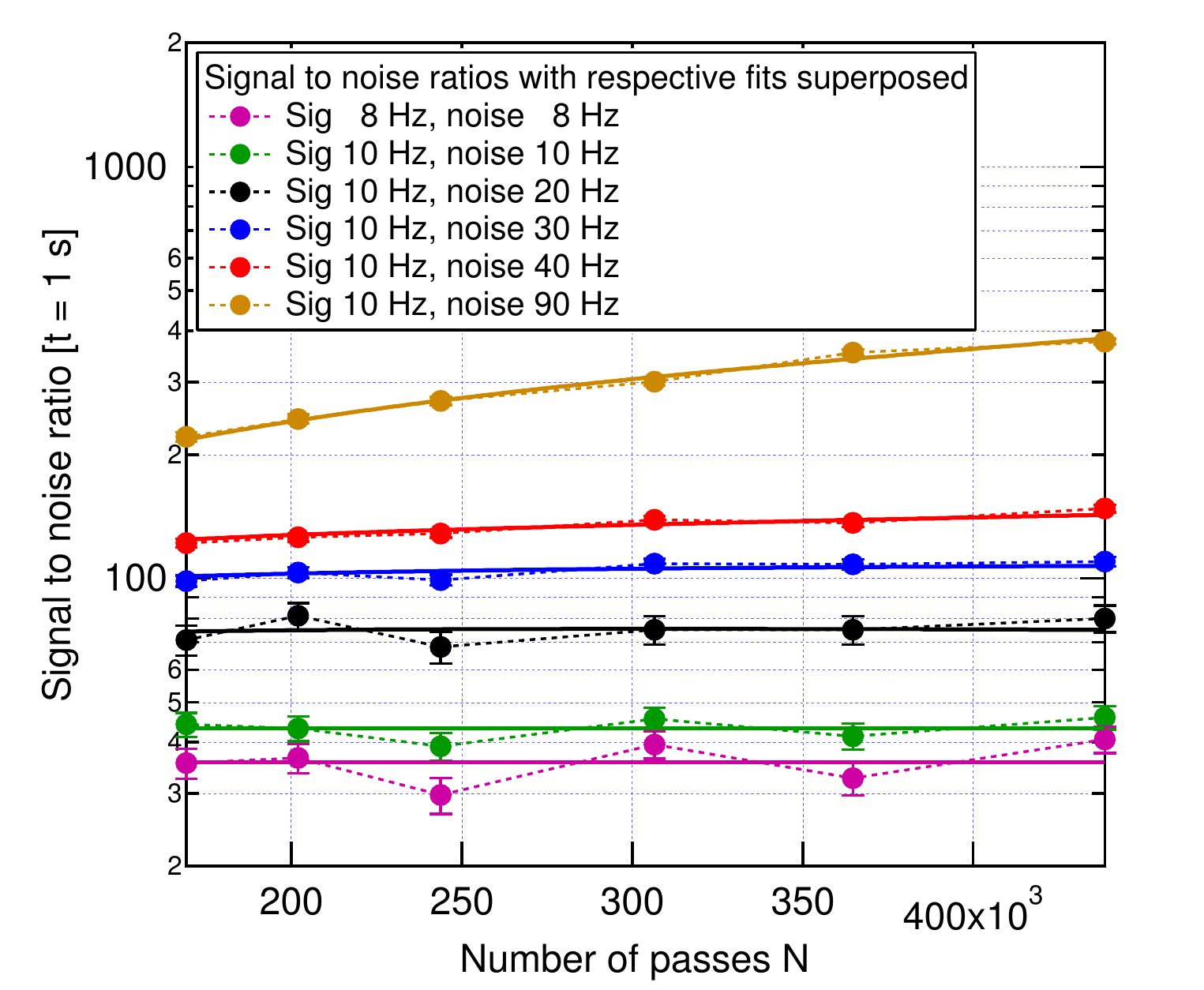}
\end{center}
\caption{signal-to-noise ratios of the Cotton-Mouton peaks with respect to the noise at different frequencies. The fits take into account the cavity response at the different frequencies according to equation (\ref{eq:snr_norm}).}
\label{fig:snr1}
\end{figure}

In figure \ref{fig:snr1} we report the signal-to-noise ratios for both the Cotton-Mouton signals at $2\nu_\alpha = 8\;$Hz (purple) and $2\nu_\beta=10\;$Hz (green) extracted from figure \ref{fig:spettri_ell}. On the same plot we have also reported the ratio of the Cotton-Mouton signal at $2\nu_\beta=10\;$Hz with respect to the noise at 20~Hz, 30~Hz, 40~Hz and 90~Hz to see whether indeed the noise is independent of finesse also at higher frequencies. The value of the noise at 8~Hz and 10~Hz is determined as the average of the noise on either side of the peaks in a frequency range of 0.5~Hz. For the other frequencies the noise is determined as the average over a 0.5~Hz frequency range. 

As can be seen, the ratios at 8~Hz and 10~Hz are indeed \emph{independent} of the finesse. The apparent increase of the signal-to-noise ratio with $N$ at higher noise frequencies 
is actually only due to the different frequency response of the cavity at the frequency of the signal and at the frequency of the noise.

Following the hypothesis that the noise in the polarimeter is dominated by an ellipticity noise \emph{per pass} $s_\psi$ we have fitted the different signal-to-noise ratios with the expression
\begin{equation}
\frac{\Psi^{\rm meas}(\nu_{\rm sig})}{S_{\Psi^{\rm meas}}(\nu_{\rm noise})}= \frac{h_{\alpha_{\rm EQ}}(\nu_{\rm sig})}{h_{\alpha_{\rm EQ}}(\nu_{\rm noise})}\frac{\psi_0}{{s_\psi(\nu_{\rm noise})}}
\label{eq:snr_norm}
\end{equation}
obtaining the superimposed fits. 

Considering a more complicated function in which one fixes a common ellipticity noise per pass $s_\psi$, a common rotation noise per pass $s_\phi$ for each value of $N$ and a flat baseline noise contribution $S_e$ according the the expression

\begin{small}
\begin{eqnarray}\nonumber
&&\frac{\Psi^{\rm meas}(\nu_{\rm sig})}{S_{\Psi^{\rm meas}}(\nu_{\rm noise})}=\\\nonumber
&&= \frac{h_{\alpha_{\rm EQ}}(\nu_{\rm sig})\psi_0}{\sqrt{s_\psi(\nu_{\rm noise})^2 +\left(\frac{N\alpha_{\rm EQ}}{2}s_\phi(\nu_{\rm noise}) h(\nu_{\rm noise})\right)^2+\frac{S_e^2}{Nk(\alpha_{\rm EQ})h_{\alpha_{\rm EQ}}(\nu_{\rm noise})}}}
\end{eqnarray}
\end{small}
does not improve the quality of the fitted data estimated using $\chi^2_{\rm ndf}$.  A global fit considering all the data in figure~\ref{fig:snr1} gives the following limits: $s_\phi/s_\psi < 0.4$ and $S_e< 2\times 10^{-8}\;$1/$\sqrt{\rm Hz}$.

The noise therefore behaves as an ellipticity noise $s_{\psi}$ generated within the cavity and multiplied by a factor $N$, just like the Cotton-Mouton signals: the total noise $S_\Psi = Ns_\psi$ is proportional to the number of passes $N$. We therefore conclude that the dominating noise source at frequencies up to $\nu = 90\;$Hz is due to a pure ellipticity noise generated in the dielectric coating of the cavity mirrors.

\begin{figure}[htb]
\begin{center}
\includegraphics[width=8.5cm]{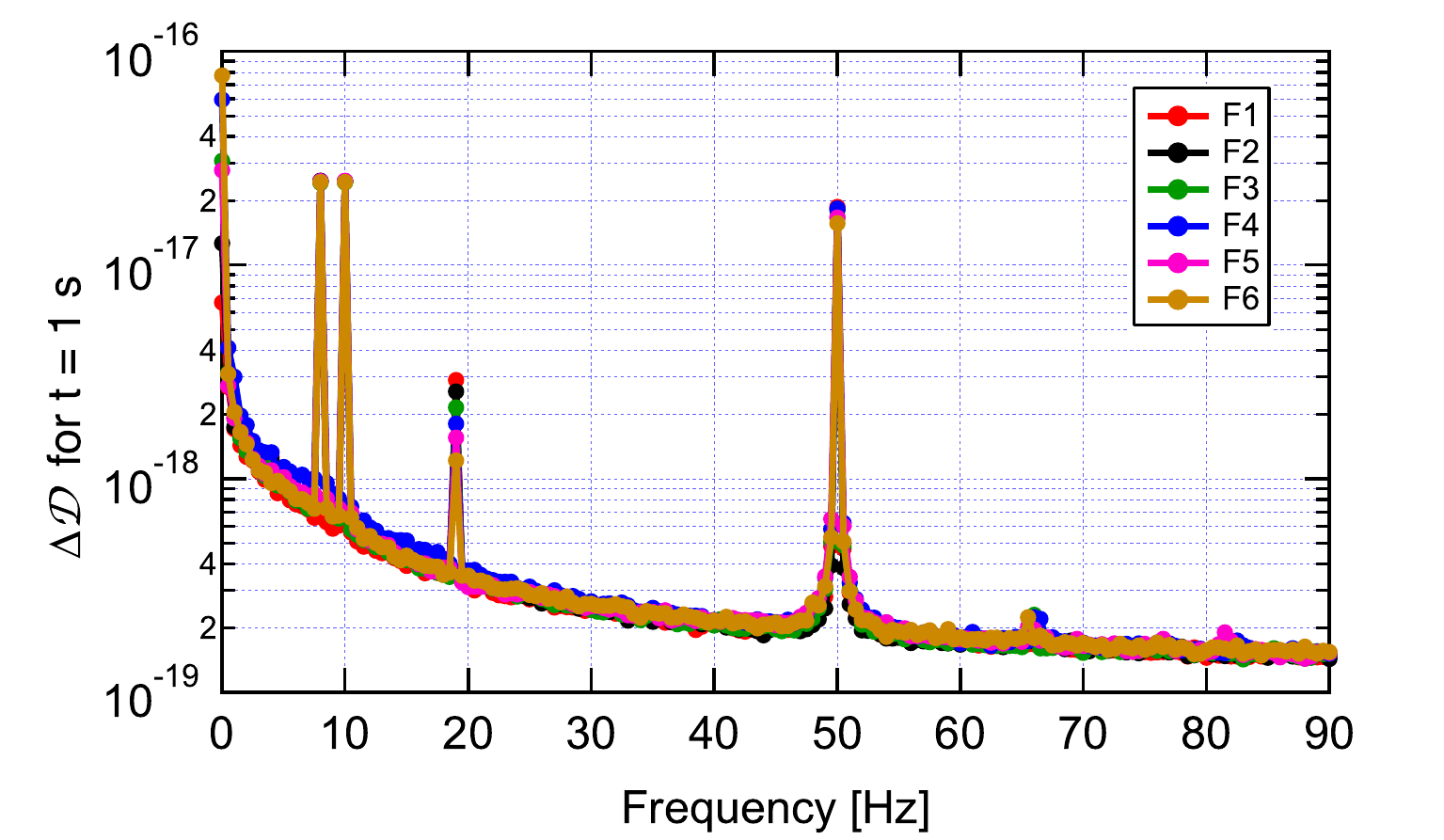}
\end{center}
\caption{The six ellipticity spectra of figure \ref{fig:spettri_ell_norm} rescaled to show the common optical path difference sensitivity independent of the number of passes $N$ assuming an ellipticity noise $S_\Psi$ proportional to $N$ and taking into account the frequency response of the cavity.}
\label{fig:spettri_opd}
\end{figure}

By rescaling figure \ref{fig:spettri_ell_norm} to obtain an optical path difference sensitivity one finds the graph in figure \ref{fig:spettri_opd}.

\subsection{Ellipticity noise versus rotation noise}
\label{rot}
\begin{figure}[htb]
\begin{center}
\includegraphics[width=8.5cm]{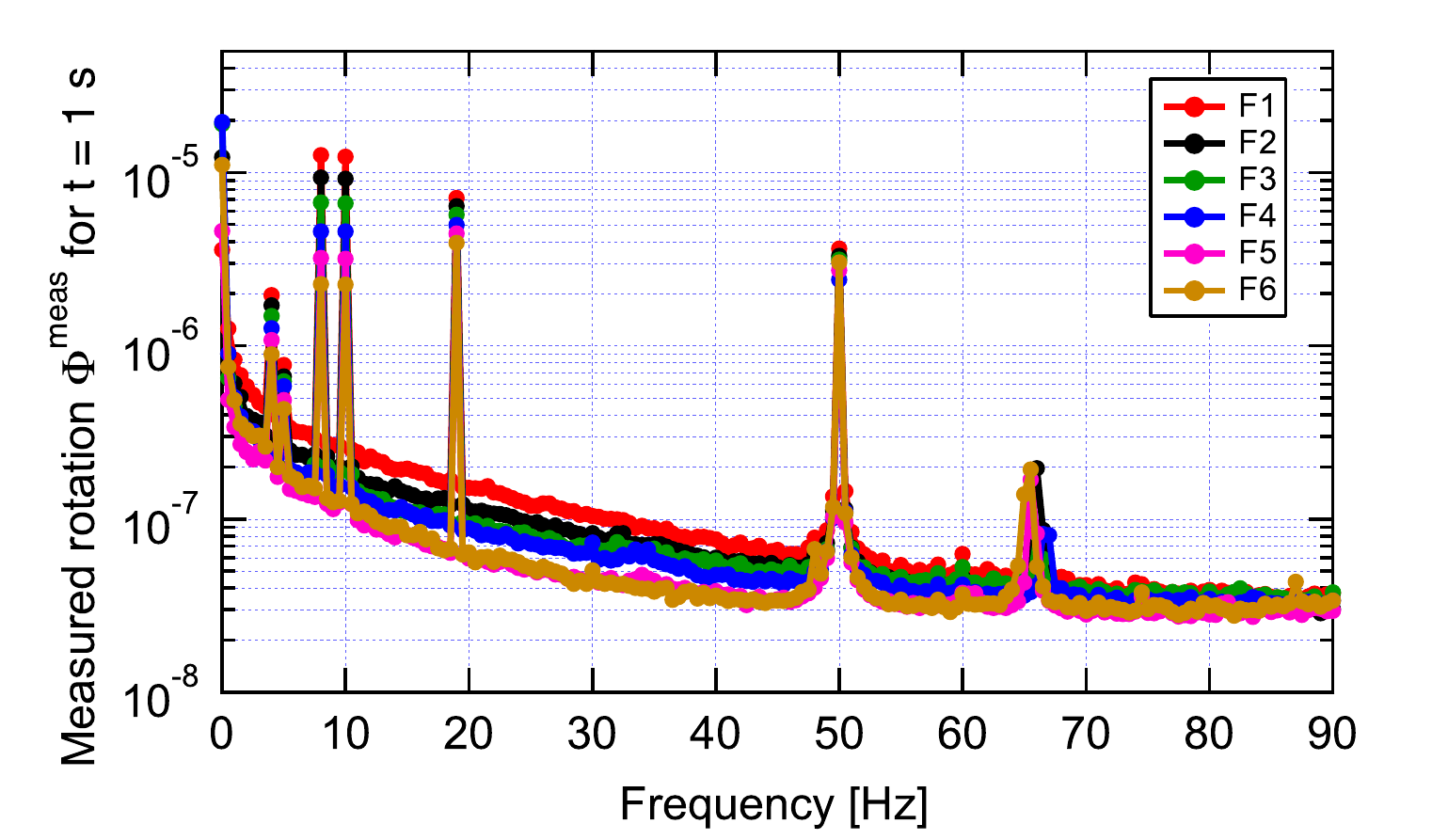}
\end{center}
\caption{Rotation spectra for the six finesse values rescaled to a 1~s integration time. The raw spectra have been rebinned by taking rms averages of the raw spectra in 0.5~Hz frequency intervals.}
\label{fig:spettri_rot}
\end{figure}

In the previous section we have discussed the ellipticity spectra for the six finesse values. In figure~\ref{fig:spettri_rot} we report the respective rotation spectra.

Two facts are apparent: the rotation noise is smaller than the ellipticity noise at all frequencies for corresponding finesse values; the noise spectra in rotation flatten above about 40~Hz at the lower finesse values. This noise floor results to be $S_r = 3.2\times 10^{-8}\;1/\sqrt{\rm Hz}$ with a dispersion around $S_r$ of $\sigma_{S_r} = 0.2\times 10^{-8}\;1/\sqrt{\rm Hz}$. This noise is slightly above the quadrature noise which corresponds to the intrinsic rotation noise of the polarimeter.

\begin{figure}[htb]
\begin{center}
\includegraphics[width=4.15cm]{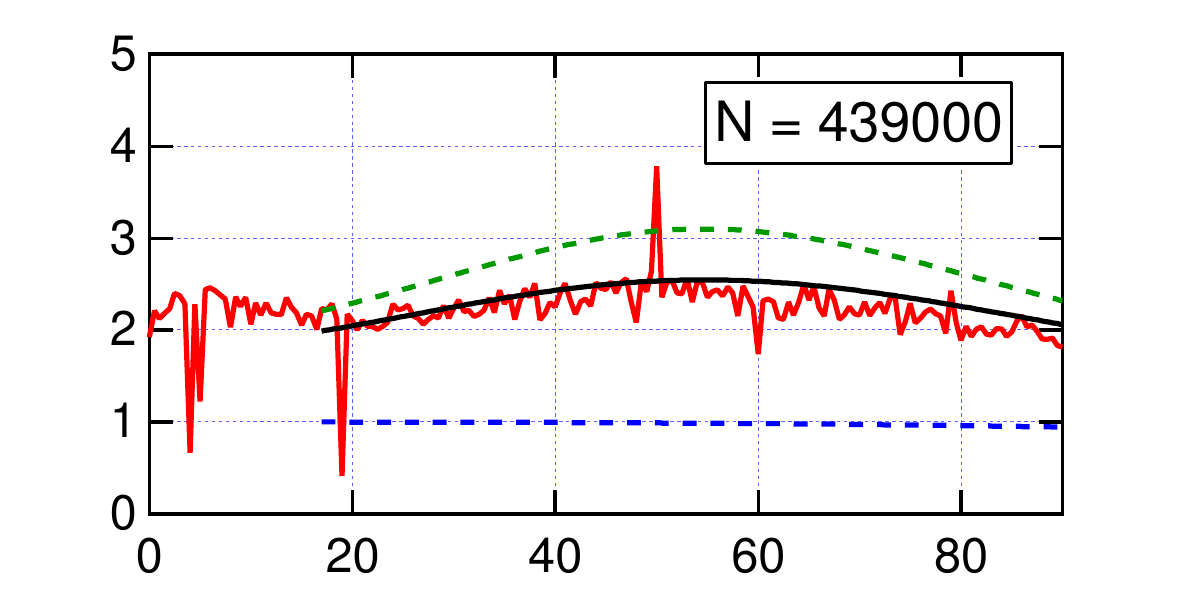}
\includegraphics[width=4.15cm]{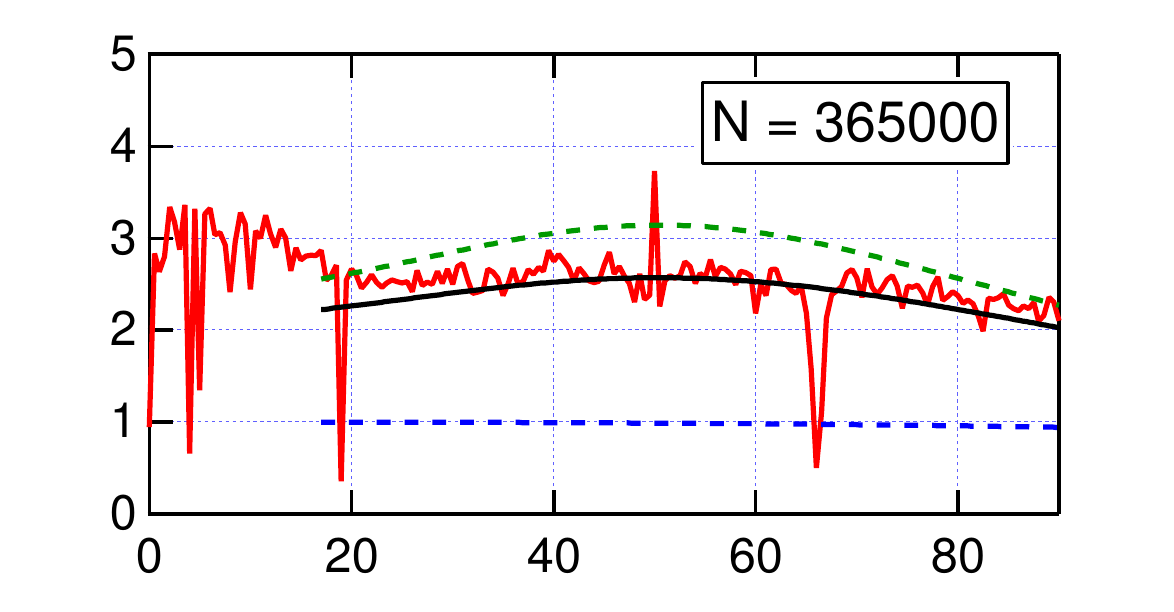}
\includegraphics[width=4.15cm]{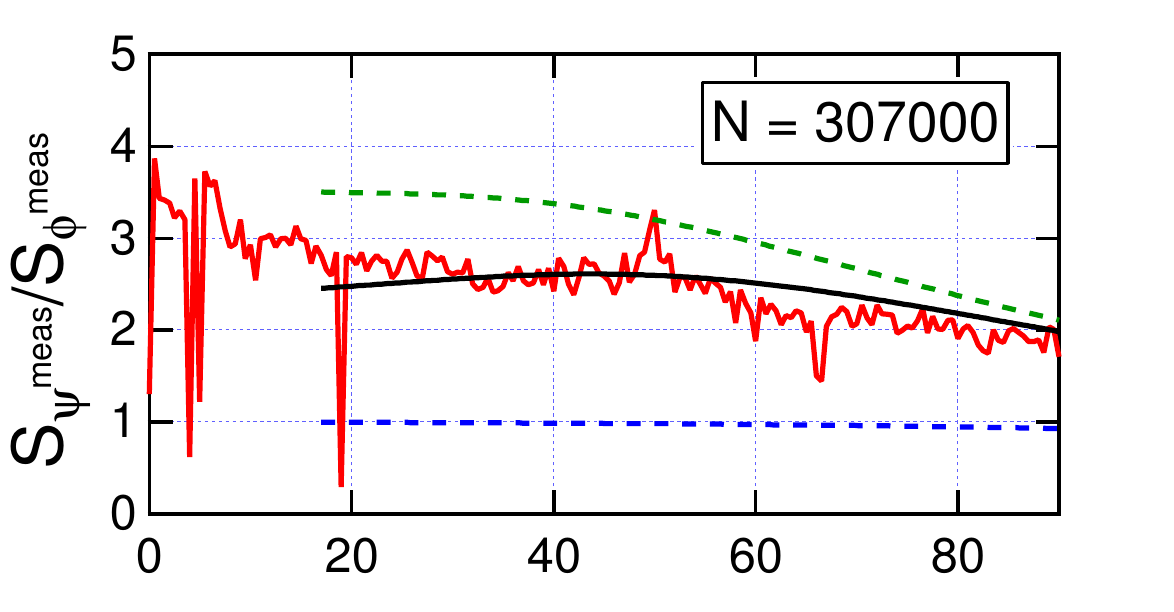}
\includegraphics[width=4.15cm]{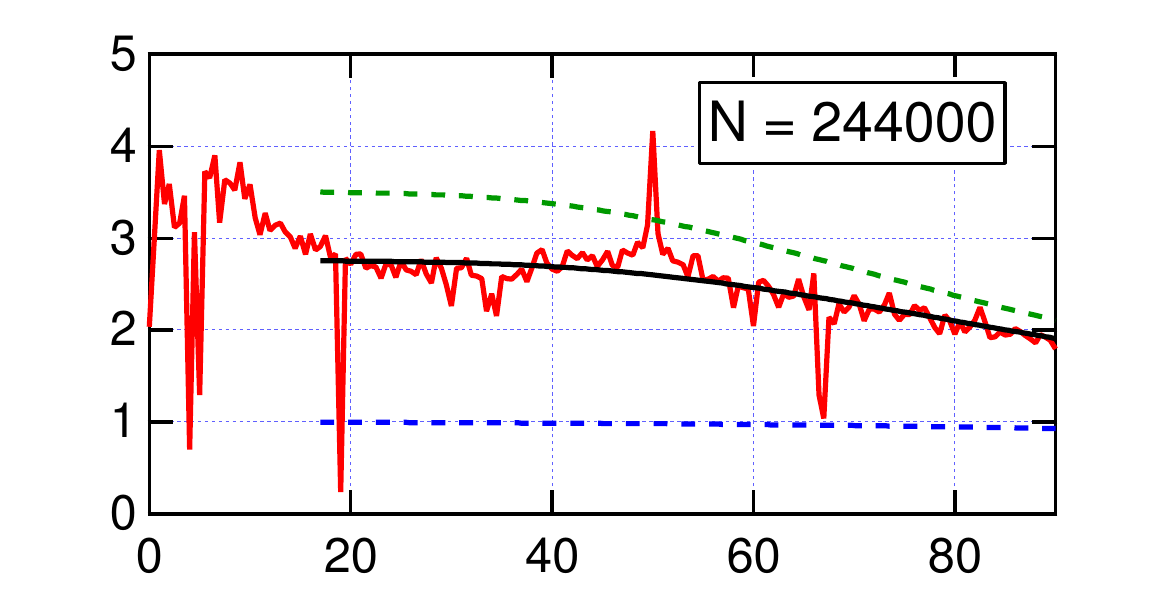}
\includegraphics[width=4.15cm]{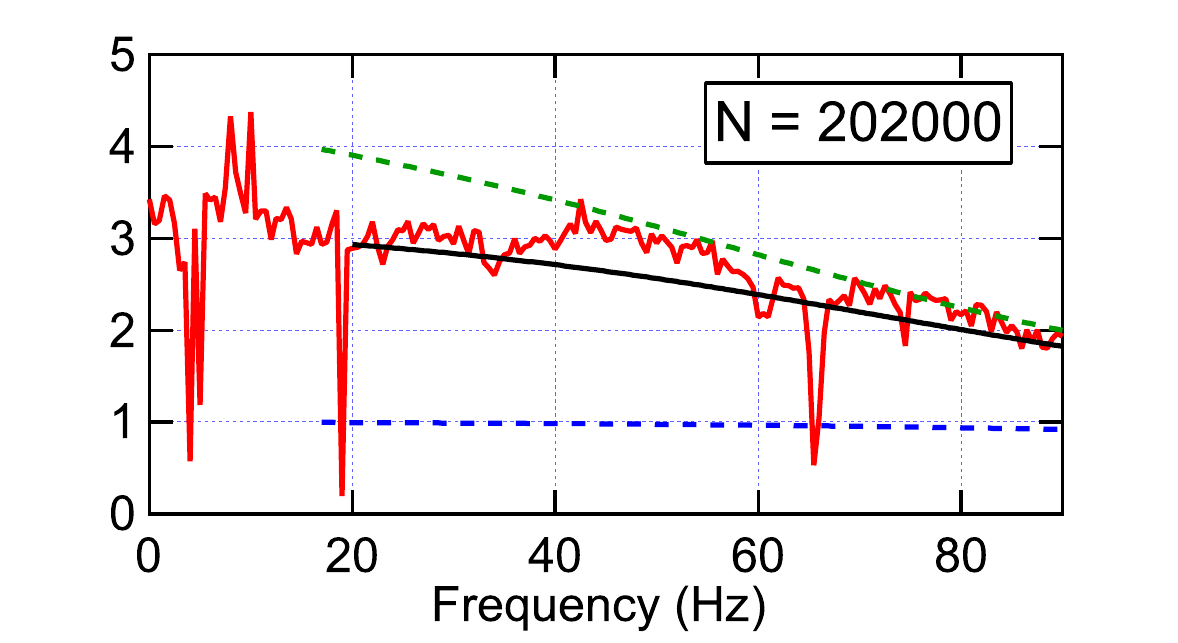}
\includegraphics[width=4.15cm]{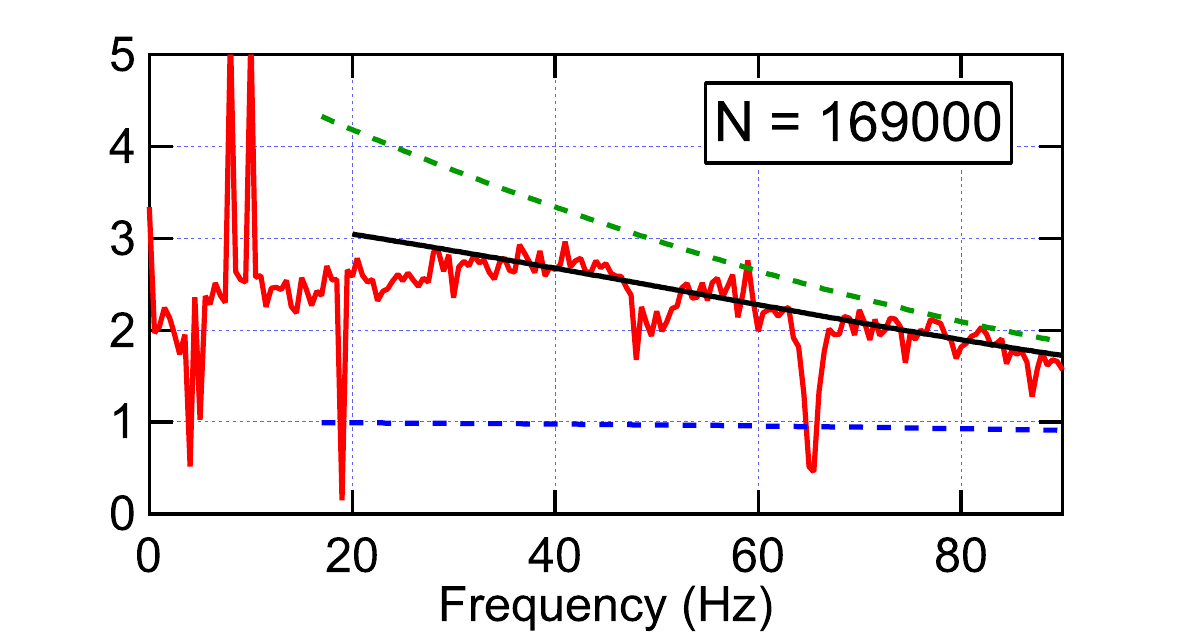}
\end{center}
\caption{Plots of the ratios of the ellipticity noise to the rotation noise. Fits are taken from 20~Hz and 90~Hz. The peak frequencies have been excluded in the fits. See text for details.}
\label{fig:ratio_ell_rot}
\end{figure}

To further confirm that the dominant noise source originates from a birefringence fluctuation we have considered the ratios of the noise spectrum in ellipticity with respect to the noise spectrum in rotation for the six different finesse values. These spectra have been fitted with the expression

\begin{eqnarray}\nonumber
&&\frac{S_{\Psi^{\rm meas}}(\nu)}{S_{\Phi^{\rm meas}}(\nu)}=\\\nonumber
&&= \frac{\sqrt{s_\psi(\nu)^2 +\left(\frac{N\alpha_{\rm EQ}}{2}s_\phi(\nu) h(\nu)\right)^2+\frac{S_e^2}{Nk(\alpha_{\rm EQ})h_{\alpha_{\rm EQ}}(\nu)}}}{\sqrt{s_\phi(\nu)^2 +\left(\frac{N\alpha_{\rm EQ}}{2}s_\psi(\nu) h(\nu)\right)^2+\frac{S_r^2}{Nk(\alpha_{\rm EQ})h_{\alpha_{\rm EQ}}(\nu)}}}
\end{eqnarray}
where we have used a frequency dependence of $s_\psi(\nu) = \sqrt{(a\nu^{-1})^2+(b\nu^{-0.25})^2}$ as a result of fitting figure \ref{fig:spettri_ell_norm}, from 10 Hz to 90 Hz, and we have assumed that the ratio $s_\phi/s_\psi$ is independent of frequency. The fit has been performed from 20~Hz to 90~Hz and the frequencies at which a peak is present have been excluded. With these assumptions we obtain the global fits shown in figure \ref{fig:ratio_ell_rot} in which the free parameters are the ratio $s_\phi/s_\psi$ and $S_e$ and the fixed value of $S_r = 3.2\times 10^{-8}\;1/\sqrt{\rm Hz}$ was used as deduced from figure \ref{fig:spettri_rot} above 40~Hz. 
The fits indicate a ratio $s_\phi/s_\psi =(0.21\pm 0.01)$ 
and $S_e \le 3 \times 10^{-8}\;1/\sqrt{\rm Hz}$ compatible with the shot noise limit shown in figure \ref{fig:sens}. 
On the same graphs we have also plotted the two cases for $s_\phi/s_\psi = 0$ (dashed green) and $s_\phi/s_\psi = 1$ (dashed blue) keeping the values for $S_e$ as obtained from the global fit and and $S_r = 3\times 10^{-8}$.

\subsection{Consequences of results}
A first important consequence of the findings presented in this section is 
that the signal-to-noise ratio in a Fabry-Perot based polarimeter with a calculated optical path difference sensitivity equal to or better than the sensitivity shown in figure \ref{fig:spettri_opd} will not improve by increasing the finesse of the cavity. Assuming a predicted shot noise sensitivity given by equation (\ref{eq:shot}), the maximum useful finesse up to which one gains in signal to noise ratio is determined by
\[
{\cal F}_{\rm max} = \sqrt{\frac{e}{I_\parallel q}}\frac{\lambda}{2 S_{\Delta{\cal D}}}
\]
where $S_{\Delta{\cal D}}$ can be read off figure \ref{fig:spettri_opd}. For the experimental configuration presented in this paper, where $S_{\Delta{\cal D}} \approx 6\times 10^{-19}\;$m/$\sqrt{\rm Hz}$ @ 10~Hz, one finds ${\cal F}_{\rm max} = 1.6\times 10^4$.

The second important fact resulting from these measurements is that the dominant source of noise is indeed due to a birefringence fluctuation in the cavity mirror coatings. 

%

\section{Noise origin}
Polarimetric measurements using a Fabry-Perot cavity to increase the effective optical path length have reached an intrinsic limit due to the coatings of the cavity mirrors. Our measurements show that this noise is due to birefringence fluctuations in the coatings which we believe are of thermal origin.

As mentioned in section \ref{mirrorbirif} cavity mirrors always present an intrinsic birefringence. There could therefore be two principal causes for these birefringence fluctuations: a fluctuation of the intrinsic birefringence; a fluctuation of the birefringence independent of the intrinsic value.
As was shown in figure \ref{fig:path}, there is a very strong correlation in the optical path difference noise between completely different experiments with very different values of ${\cal F}$. This seems to indicate that the source of the intrinsic birefringence noise is independent of the intrinsic mirror birefringence inducing the retardations $\alpha_{1,2}$. 

We also note that any polarization effect intrinsic to the cavity, be it static or dynamical, is generated in the first reflecting layers encountered by the light from inside the cavity.
With a transmittance of the mirrors $T=2.4\times10^{-6}$, as is the case of the PVLAS experiment \cite{OE2014}, the electric field inside the reflective coatings has an exponential decay with
\[
\frac{N_{\rm LP}}{\lambda_{\rm LP}}=-\ln\sqrt{T}
\]
where $N_{\rm LP}\approx 20$ is the total number of high refractive index - low refractive index pairs composing the reflective coating and $\lambda_{\rm LP}$ represents the number of coating pairs after which the electric field (as opposed to the intensity) has decreased to $1/e$  of the incident field. One finds $\lambda_{\rm LP} = 3.0$. Most of the ellipticity signals is therefore accumulated in the first $\lambda_{\rm LP}$ pairs of dielectric coatings for each reflection. 
This corresponds to a geometrical thickness $d_{\rm LP} \simeq 1\;\mu$m. These considerations further justify the use of an extra loss located outside the mirrors as a means to study the intrinsic birefringence noise of the mirrors.

There are three possible causes for birefringence thermal noise in a medium: direct temperature dependence of the index of refraction (thermo-refractive effect); indirect temperature dependence of the index of refraction due to a linear expansion coefficient coupled to a stress optic coefficient; volume fluctuations to Brownian motion. Here we will discuss only the two first effects.  
\subsection{Thermo-refractive effect}

Let us consider the index of refraction of the mirror coatings along the natural axes of a mirror as $n_\parallel$ and $n_\perp$ resulting in a birefringence $\Delta n = n_\parallel - n_\perp$. The optical path through the coating of a mirror per reflection for light polarised parallel and perpendicularly to the slow axis (here considered to be the $\parallel$ direction) will be
\begin{eqnarray}
{\cal D}_\parallel &\approx& 2\int\limits_{d_{\rm LP}} n_\parallel\;dl\\
{\cal D}_\perp &\approx& 2\int\limits_{d_{\rm LP}} n_\perp\;dl
\end{eqnarray}
where the factor 2 is intended to take into account the round trip inside the coating. The intrinsic optical path difference of a mirror coating per reflection will therefore be 
\[
\Delta{\cal D} = {\cal D}_\parallel - {\cal D}_\perp \approx 2\int\limits_{d_{\rm LP}} \Delta n\;dl.
\]

By considering the thermo-refractive effect due to a temperature dependence of $n = n({\rm T})$, 
the optical path difference temperature dependence will be
\begin{eqnarray}
\frac{d\Delta{\cal D}({\rm T})}{d{\rm T}} \approx 2\int\limits_{d_{\rm LP}} \frac{d\Delta n({\rm T})}{d{\rm T}}\;dl.
\end{eqnarray}
Hence ${d\Delta{\cal D}({\rm T})}/{d{\rm T}}\ne0$ only if 
$d n_\parallel /d{\rm T}\ne d n_\perp /d{\rm T}$. In this case the optical path difference spectral density $S_{\Delta {\cal D}}(\nu)$ due to the thermo-refractive effect 
will be
 \[
S_{\Delta {\cal D}}(\nu) = \frac{d\Delta{\cal D}({\rm T})}{d{\rm T}}S_{\rm T}(\nu)
 \]
where $S_{\rm T}(\nu)$ is the temperature noise spectral density.

An estimate of $\Delta {\cal D}$, due to the intrinsic mirror birefringence, can be obtained from the value of $\alpha_{\rm EQ} = 2.3\times 10^{-6}$, is
\[
\Delta {\cal D} \approx 2\int\limits_{d_{\rm LP}}{\Delta n \;dl} = \frac{\alpha_{\rm EQ}}{2\pi}\lambda \approx 4\times10^{-13}\;{\rm m}.
\]
A rough value for $\frac{1}{\Delta n}\frac{d\Delta n}{d{\rm T}} \sim 10^{-5}\;$K$^{-1}$ for fused silica can be deduced from the expressions reported in \cite{tan} considering $n_\parallel \simeq n_\perp$. Therefore
\[
\frac{d\Delta {\cal D}}{d{\rm T}} = 2\int\limits_{d_{\rm LP}}{\frac{d\Delta n}{d{\rm T}}\;dl} \sim 10^{-5}\times2\int\limits_{d_{\rm LP}}{\Delta n \;dl}\sim 4\times 10^{-18}\;\frac{\rm m}{\rm K}.
\]
Following \cite{braginsky2} the temperature fluctuations averaged over a volume $\pi r_0^2 d_{\rm LP}/2$ occupied by the Gaussian power profile of waist $r_0$ being reflected using a weight function $q(\vec r)$
\[
q(\vec{r}) = \frac{2}{\pi r_0^2 d_{\rm LP}}e^{-\left(x^2 + y^2\right)/r_0^2}\; e^{-2z/d_{\rm LP}}
\]
results in a temperature noise spectral density $S_{\rm T}(\nu)$ \cite{braginsky2}
\begin{equation}
S_{\rm T}^2(\nu) = {\frac{\sqrt{2}\kappa_B{\rm T}^2}{\pi r^2_0\sqrt{2\pi\nu\rho C_{\rm T}  \lambda_{\rm T}}}} = \frac{\sqrt{2}k_B{\rm T}^2}{2\pi\nu\rho C_{\rm T} r_{\rm T}^3}\frac{r_{\rm T}^2}{\pi r_0^2}
\label{ST}
\end{equation}
where $\kappa_B$ is the Boltzmann constant, $\rho$ is the density, $C_{\rm T}$ is the specific heat capacity, $\lambda_{\rm T}$ is the thermal conductivity and
\[
r_{\rm T} = \sqrt{\frac{\lambda_{\rm T}}{\rho C_{\rm T} 2\pi\nu}}
\]
is the characteristic diffuse heat transfer length ($d_{\rm LP}\ll r_{\rm T}\ll r_0$). 
Considering fused silica (FS), for which $\rho = 2200\;$kg/m$^3$, $C_{\rm T} = 670\;$J/(kg K) and $\lambda_{\rm T} = 1.4\;$W/(m K) this results in
\[
S^{\rm FS}_{\rm T}(\nu) \simeq 2\times 10^{-8}\;\frac{{\rm K}}{\sqrt{\rm Hz}}\;@\;1\;{\rm Hz}
\]
having set the beam diameter $r_0 \approx 0.5\;$mm. Considering tantala, Ta$_2$O$_5$, (TA) instead (we are assuming this is the material used for the high-index layer in the mirror coating), for which $\rho = 8200\;$kg/m$^3$, $C_{\rm T} = 300\;$J/(kg K) and $\lambda_{\rm T} = \left[0.026  \div 15\right]\;$W/(m K) (for a film) \cite{lambdaLP}, one finds
\[
S^{\rm TA}_{\rm T}(\nu) \simeq (1\div6)\times 10^{-8}\;\frac{{\rm K}}{\sqrt{\rm Hz}}\;@\;1\;{\rm Hz}.
\]


With the above values the thermo-refractive noise spectral density in optical path difference $S^{\rm TR}_{\Delta {\cal D}}({\nu})$ can be estimated to be of the order
\[
S^{\rm TR}_{\Delta {\cal D}}({\nu}) = \frac{d\Delta {\cal D}}{d{\rm T}}\sqrt{\frac{\sqrt{2}k_B{\rm T}^2r_{\rm T}(\nu)}{\pi r_0^2 \lambda_{\rm T}}}\sim  10^{-25}\;\frac{\rm m}{\sqrt{\rm Hz}}\;@\;1\;{\rm Hz}
\]
well below the measured values reported in figure \ref{fig:path} and figure \ref{fig:spettri_opd}. We therefore believe that the source of noise in the PVLAS polarimeter is not due to a thermo-refractive effect.
\subsection{Stress induced birefringence}

Length fluctuations 
will generate birefringence through the stress optic coefficient. Indeed given a stress optical coefficiente $C_{\rm SO}$ and a Young's modulus $Y$ the induced variation in the index of refraction due to stress is given by
\begin{equation}
\delta n_{\parallel,\perp} = C_{\rm SO} Y \left(\frac{\delta l_{\parallel,\perp}}{l}\right)
\end{equation}
where ${\delta l_{\parallel,\perp}}/{l}$ is a relative length variation along two perpendicular directions $\parallel$ and $\perp$ over a length $l$. Again following the considerations in \cite{braginsky2} an order of magnitude estimate of the induced birefringence noise spectral density $S_{\Delta n}(\nu)$ over the spot size of the reflected beam due to temperature fluctuations related to stress can be made. 

From equation (\ref{ST}) for $S_{\rm T}(\nu)$, the averaged relative length variations over a length $r_{\rm T}$, indicated by the brackets $\langle\rangle_{\parallel,\perp}$, along two perpendicular directions $\parallel$ and $\perp$ will be
\[
\left\langle\frac{{\delta r_{\rm T}}}{r_{\rm T}}\right\rangle_{\parallel,\perp} = \alpha_{\rm T} S_{\rm T}(\nu) = \alpha_{\rm T}\sqrt{\frac{\sqrt{2}\kappa_B{\rm T}^2}{\pi r^2_0\sqrt{2\pi\nu\rho C_{\rm T}  \lambda_{\rm T}}}}
\]
where $\alpha_{\rm T}$ is the linear expansion coefficient. This will generate
a birefringence noise spectral density
\begin{eqnarray}
S_{\Delta n}(\nu) &\simeq& C_{\rm SO} Y \sqrt{\left\langle\frac{\delta r_{\rm T}}{r_{\rm T}}\right\rangle_\parallel^2+\left\langle\frac{\delta r_{\rm T}}{r_{\rm T}}\right\rangle_\perp^2}=\nonumber\\
&=& C_{\rm SO} Y \alpha_{\rm T}\sqrt{\frac{2\sqrt{2}\kappa_B{\rm T}^2}{\pi r^2_0\sqrt{2\pi\nu\rho C_{\rm T}  \lambda_{\rm T}}}}.
\end{eqnarray}

A very rough estimate of the optical path difference spectral density noise $S_{\Delta{\cal D}}=2\int_{d_{\rm LP}}{S_{\Delta n} \;dl}$ accumulated in a reflection results in
\begin{eqnarray}
S_{\Delta{\cal D}} \approx 2S_{\Delta n}d_{\rm LP} =C_{\rm SO} Y \frac{\alpha_{\rm T} d_{\rm LP}}{r_{0}}\sqrt{\frac{8{\rm T}^2\kappa_B}{\pi \sqrt{\pi\nu\rho C_{\rm T}  \lambda_{\rm T}}}}.
\end{eqnarray}
For fused silica for which $\alpha_{\rm T} = 5\times 10^{-7}\;$K$^{-1}$, $Y = 70\;$GPa and $C_{\rm SO} = 3\times 10^{-12}\;$Pa$^{-1}$ one finds
\[
S^{\rm FS}_{\Delta{\cal D}}\sim 7\times 10^{-21}\;\frac{{\rm m}}{\sqrt{\rm Hz}}\;@\;1\;{\rm Hz} \nonumber
\]
whereas for tantala 
\[
S^{\rm TA}_{\Delta{\cal D}}\sim (1\div6)\times 10^{-19}\;\frac{{\rm m}}{\sqrt{\rm Hz}}\;@\;1\;{\rm Hz} \nonumber
\]
where the values for tantala are $Y= 150\;$GPa and $\alpha_{\rm T} = 8\times 10^{-6}\;$K$^{-1}$ and we have use $C_{{\rm SO}} = 3\times 10^{-12}\;$Pa$^{-1}$ for fused silica not having found a value for tantala in the literature.

The value for $S^{\rm TA}_{\Delta{\cal D}}$ in the case of tantala is quite close to the measured values especially at higher frequencies. 
The exact expression for $S_{\Delta {\cal D}}$ is beyond the scope of this paper but indeed a stress mechanism could generate a birefringence noise of the same order of magnitude as the one measured.

This stress will be present both in the substrate and in the mirror coatings. As discussed above, given that the electric field within the coating is strongest in the first $\lambda_{\rm LP}$ layers encountered by the light in the cavity, the induced $S_{\Delta {\cal D}}$ will be dominated by these first layers and in particular by the tantala layers. 
\section{Conclusions}
Birefringence noise limits the sensitivity of precision measurements in ellipsometers like those designed to detect the birefringence of vacuum due to  magnetic fields. We have measured the noise present in the PVLAS polarimeter in both ellipticity and rotation modes along with Cotton-Mouton and Faraday signals as a function of the finesse of the Fabry-Perot cavity. We have shown that the signal-to-noise ratio of the Cotton-Mouton ellipticity signals
is {\em independent} of the finesse of the cavity as is the ellipticity noise. We have shown that for the rotation noise this is not the case. We have also studied the ellipticity noise to rotation noise ratios which confirm that the dominant noise source in the polarimeter is due to a fluctuating birefringence {\em inside} the Fabry-Perot cavity.

We infer that the noise is generated in the first few layers of the mirror coatings and that the origin is due to thermally induced stress fluctuations namely due to a thermo-elastic effect.

It is therefore apparent that the continuous search to improve the sensitivity in optical path difference $S_{\Delta {\cal D}}$ by increasing the finesse of the Fabry-Perot cavity has reached a limit.

The quest to measure vacuum magnetic birefringence using optical techniques must therefore 
\begin{itemize}
\item{reduce the optical path difference noise by cooling the mirrors and/or by finding new materials for the coatings with a lower stress optic coefficient or lower linear expansion coefficient;}
\item{decrease the number of reflections (finesse) and increase the cavity and magnetic field lengths to preserve the optical path length.}
\item{increase the vacuum magnetic birefringence signal using high, long, static superconducting fields and inducing the necessary signal modulating for improved sensitivity by varying the polarisation \cite{PolSchem}.}
\end{itemize}

Finally let us note that if the intrinsic total retardation $\frac{N\alpha_{\rm EQ}}{2}$ can be kept low in such a way that the mixing between ellipticity and rotation is also small, then increasing the finesse during rotation measurements is advantageous. Indeed given that the dominant source of ellipticity noise is due to birefringence fluctuations, rotation sensitivity may not be limited by an intrinsic thermal source. This could lead to improved laboratory experimental limits on the existence of axion like particles \cite{DellaValle2015,Aldo,MPZ,ni}.
%
%
%

\end{document}